\documentclass[%
 reprint,
superscriptaddress,
longbibliography,
 amsmath,amssymb,
 aps,
prb,
]{revtex4-1}
\setcitestyle{numbers,square}

\usepackage{graphicx}
\usepackage{dcolumn}
\usepackage{bm}
\usepackage{amsfonts}
\usepackage{xcolor}
\usepackage{enumerate}
\usepackage[normalem]{ulem}
\usepackage{color}

\usepackage{hyperref}
\hypersetup{
    colorlinks,%
    citecolor=blue,%
    linkcolor=blue,%
    urlcolor=blue
}

\newcommand{\orcid}[1]{\href{https://orcid.org/#1}{\includegraphics[width=8pt]{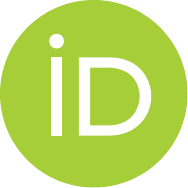}}}

\newcommand{\ncl}{nCL}
\newcommand{\nclphi}{nCL$^{\rm phil}$}
\newcommand{\nclpho}{nCL$^{\rm phob}$}
\newcommand{\nclcl}{nCL$_2$}
\newcommand{\nclclphi}{nCL$_2^{\rm phil}$}
\newcommand{\nclclpho}{nCL$_2^{\rm phob}$}
\newcommand{\nclclclphi}{nCL$_3^{\rm phil}$}
\newcommand{\nclclclpho}{nCL$_3^{\rm phob}$}

\begin{document}

\title{Nanoscale Structural and Electronic Properties of Cellulose/Graphene Interfaces}

\author{Gustavo H. Silvestre\orcid{0000-0002-4010-2656}}
\affiliation{Instituto de F\'isica, Universidade Federal de Uberl\^andia, 38400-902, Uberl\^andia, MG, Brazil}

\author{Felipe Crasto de Lima\orcid{0000-0002-2937-2620}}
\affiliation{Ilum School of Science, Brazilian Center for Research in Energy and Materials (CNPEM), Campinas, SP, 13083-970, Brazil}

\author{Juliana S. Bernardes\orcid{0000-0002-2758-0880}}
\affiliation{Brazilian Nanotechnology National Laboratory, Brazilian Center for Research in Energy and Materials (CNPEM), Campinas, SP, 13083-970, Brazil}
\affiliation{Center for Natural and Human Sciences, Federal University of ABC, Santo Andr\'e, S\~ao Paulo 09210-580, Brazil}

\author{Adalberto Fazzio\orcid{0000-0001-5384-7676}}
\affiliation{Ilum School of Science, Brazilian Center for Research in Energy and Materials (CNPEM), Campinas, SP, 13083-970, Brazil}
\affiliation{Center for Natural and Human Sciences, Federal University of ABC, Santo Andr\'e, S\~ao Paulo 09210-580, Brazil}

\author{Roberto H. Miwa\orcid{0000-0002-1237-1525}}
\affiliation{Instituto de F\'isica, Universidade Federal de Uberl\^andia, 38400-902, Uberl\^andia, MG, Brazil}

\date{\today}

\begin{abstract}

The development of electronic devices based on the functionalization of (nano)cellulose platforms relies upon an atomistic understanding of the structural, and electronic properties of the combined system, cellulose/functional element.  In this work, we present a theoretical study of the nanocellulose/graphene interface (\ncl/G) based on first-principles calculations. We find that the binding energies of both hydrophobic/G (\nclpho/G) and hydrophilic/G (\nclphi/G) interfaces are primarily dictated by the van der Waals interactions, and are comparable with that of their 2D interface counterparts. We verify that the energetic preference of \nclpho/G has been reinforced by the inclusion of an aqueous media via the implicit solvation model. Further structural characterization was carried out using a set of simulations of Carbon K-edge X-ray absorption spectra to identify and distinguish the key absorption features of the \nclpho/G and \nclphi/G interfaces. The electronic structure calculations reveal that the linear energy bands of graphene lie in the band gap of the  \ncl\, sheet, while depletion/accumulation charge density regions are observed. We show that external agents, i.e. electric field and mechanical strain, allow for tunability of the Dirac cone and the charge density at the interface. The control/maintenance of the Dirac cone states in \ncl/G is an important feature for the development of electronic devices based on cellulosic platforms.

\end{abstract}

\maketitle

\section{Introduction}

Designing new biodegradable electronic devices based on renewable and environmentally sustainable platforms has been intensively investigated with the motivation to combat resource constraints and waste disposal challenges \cite{irimia2014green}. Most electronics are still assembled from nonrenewable and nonbiodegradable materials and occasionally use production techniques that rely on hazardous compounds.

The possibility of building flexible devices using paper has led to the development of novel green electronic alternatives \cite{zhu2016wood,yao2017paper,zhao2021cellulose}. Cellulosic substrates have been explored for many applications, including transistors \cite{conti2020low}, supercapacitors \cite{gui2013natural}, and organic solar cells \cite{brunetti2019printed}. More recently, nanoparticles extracted from cellulose pulps (cellulose nanocrystals-CNC and cellulose nanofibers-CNF) have also been considered lightweight and robust materials for electronic devices \cite{du2017nanocellulose, hoeng2016use}. Substrates produced from CNCs and CNFs display advantages over regular paper, including smoothness, high optical transparency, and superior mechanical properties \cite{moon2011cellulose, salas2014nanocellulose}. Besides, due to relatively inexpensive isolation methods, nanocellulose has excellent potential as a sustainable nanomaterial for designing many functional structures. 

To impart electrical conductivity to cellulose, metallic particles \cite{inui2015miniaturized}, conductive polymers \cite{yang2015cellulose}, carbon-based particles \cite{fingolo2021enhanced}, and 2D materials \cite{cao2019solution} are usually integrated into nanocellulose through different techniques (coating, dipping, printing, blending, etc.). The combination of nanocellulose and 2D nanomaterials such as graphene, MoS$_2$, and MXene has recently triggered great interest in the scientific community as a new class of multifunctional hybrid compounds. By assembling graphene and nanocellulose within a stretchable elastomer matrix, Weng et al. \cite{weng2011graphene} fabricated a robust strain sensor for efficient human-motion detection. Moisture-responsive foldable actuators were also produced from exfoliated graphene and amphiphilic nanocellulose by a simple vacuum filtration method \cite{xu2019aqueous}. Tian et al. \cite{tian2019multifunctional} combined the excellent mechanical properties of CNF with the metal-like electrical conductivity of MXenes to design supercapacitor electrodes with high electronic conductivity of 2.95 $\times$ 10$^4$ S m$^{-1}$.

For electronic applications, 2D/nanocellulose hybrid materials should be able to tolerate mechanical stress and deformations while maintaining the satisfactorily electrical conductivity of 2D materials. Therefore, fundamental understanding of how the insertion of cellulose, a dielectric compound, influences the electrical properties of 2D materials is essential to guide the development of (nano) devices. Considerable experimental works on 2D/nanocellulose hybrids have shown promising outcomes. However, they only focus on the global electrical response and ignore the effect of nanocellulose/2D interaction at the nano and atomic levels on the electronic properties.
Our group recently used first-principles calculations with a machine learning approach to evaluate relevant chemical and structural parameters that govern the binding energy of graphene oxide/nanocellulose interfaces \cite{petry2022machine}, which have been considered promising polymeric composites for gas barriers \cite{mianehrow2020strong} and water decontamination \cite{zhu2017self}. In the same direction, Zhu et al. \cite{zhu2022revealing} employed first-principles methods to study the interface bonding behavior of graphene oxide and cellulose derivatives composite systems. {Despite such previous studies, a detailed picture of the interaction between the different cellulose surfaces and graphene is hitherto unexplored.}

In the current study, based on first-principles calculations, we investigate the (i) energetic stability, (ii) structural, and (iii) electronic properties of the nanocellulose/graphene interface (nCL/G). The hydrophobic (\nclpho/G) and hydrophilic (\nclphi/G) interfaces were addressed in this study. In (i), we have examined the role played by the vdW interaction on the nanocellulose - graphene binding strength and the aqueous media effect on the nCL/G binding energy. The structural characterization [(ii)] was performed by a set of X-ray near edge absorption structure (XANES) simulations of the \nclpho/G and \nclphi/G interfaces to identify the corresponding X-ray fingerprints. In (iii), we show that the Dirac bands of graphene lie in the nCL’s bandgap, with the Dirac point at about 2 eV above the valence band maximum of the nanocellulose ($\Delta E_\text{DP}$), followed by a small net charge transfer ($\Delta\rho\propto 10^{12} e$/cm$^2$) from graphene to the nCL/G interface region. In the sequence, we investigate the role of external agents such as electric field and mechanical strain in the electronic properties of nCL/G.

\section{Computational Details}

All calculations were performed within the density functional theory (DFT), where the exchange-correlation term was described by the generalized gradient approximation (GGA-PBE) \cite{PBE} proposed by Perdew, Burke and Ernzerhof. The periodic image interaction was avoided by using at least 25 \AA~ vacuum perpendicular to graphene sheet. The Kohn-Sham orbitals were expanded in a plane wave basis set with an energy cutoff of 400 eV and the electron-ion interaction have been evaluated using the PAW (projected augmented wave) method \cite{paw}. The Brillouin Zone (BZ) sampling was performed according to the Monkhorst-Pack scheme \cite{mp}, using a 6$\times$6$\times$1 mesh. The search for binding energies, equilibrium geometry and electronic properties were performed using Vienna Ab-initio Simulation Package (VASP) \cite{vasp1,vasp2}, and the influence of an aqueous environment was simulated based on the implicit solvation model implemented in DFT code VASP (VASPsol \cite{VASPsol-Software, mathew2014implicit, mathew2019implicit}). The equilibrium configuration was calculated taken in account a fully relaxed  of atomic positions, considering a convergence criteria of 25\,meV/\AA\, for the atomic forces. In order to provide a more complete picture of the energetic features of the \ncl/G interfaces, we have examined  role played by the vdW dispersion interaction on the \ncl/G binding energies. We have considered two different non-local vdW approaches, viz.: vdW-DF \cite{thonhauserPRL2015, thonhauserPRB2007, langrethJPhysC2009}, and vdW-optB86b \cite{klimesJPhysC2010, klimevsPRB2011}.

The Carbon K-edge X-ray absorption near edge structure (XANES) spectra were simulated using XSpectra package \cite{xas1, xas2, xas3}, implemented in Quantum ESPRESSO \cite{espresso, berland2015van}. To describe the K-edge spectra, we used a reconstructed ultrasoft pseudopotential with a core-hole in C-$1s$ orbital and the electron wave functions were recovered using GIPAW \cite{gipaw} reconstruction. Here, the BZ sampling was the same previously described and the energy cutoffs for the plane wave basis set and self-consistent total charge density were respectively 48 and 192 Ry.

\section{Results}

The structural models of \ncl/G interface are depicted in Fig.\,\ref{models}. The graphene layer interacts with the hydrophobic and hydrophilic nanocellulose sheets described by a single layer of cellulose fibrils, labelled as \nclpho/G and \nclphi/G in Fig.\,\ref{models}(a1) and (b1), and bilayer of cellulose fibrils, \nclclpho/G and \nclclphi/G in Fig.\,\ref{models}(a2) and (b2). 

\subsection{\ncl/G Binding Energy and Geometry} 
\begin{figure}
\includegraphics[width=\columnwidth]{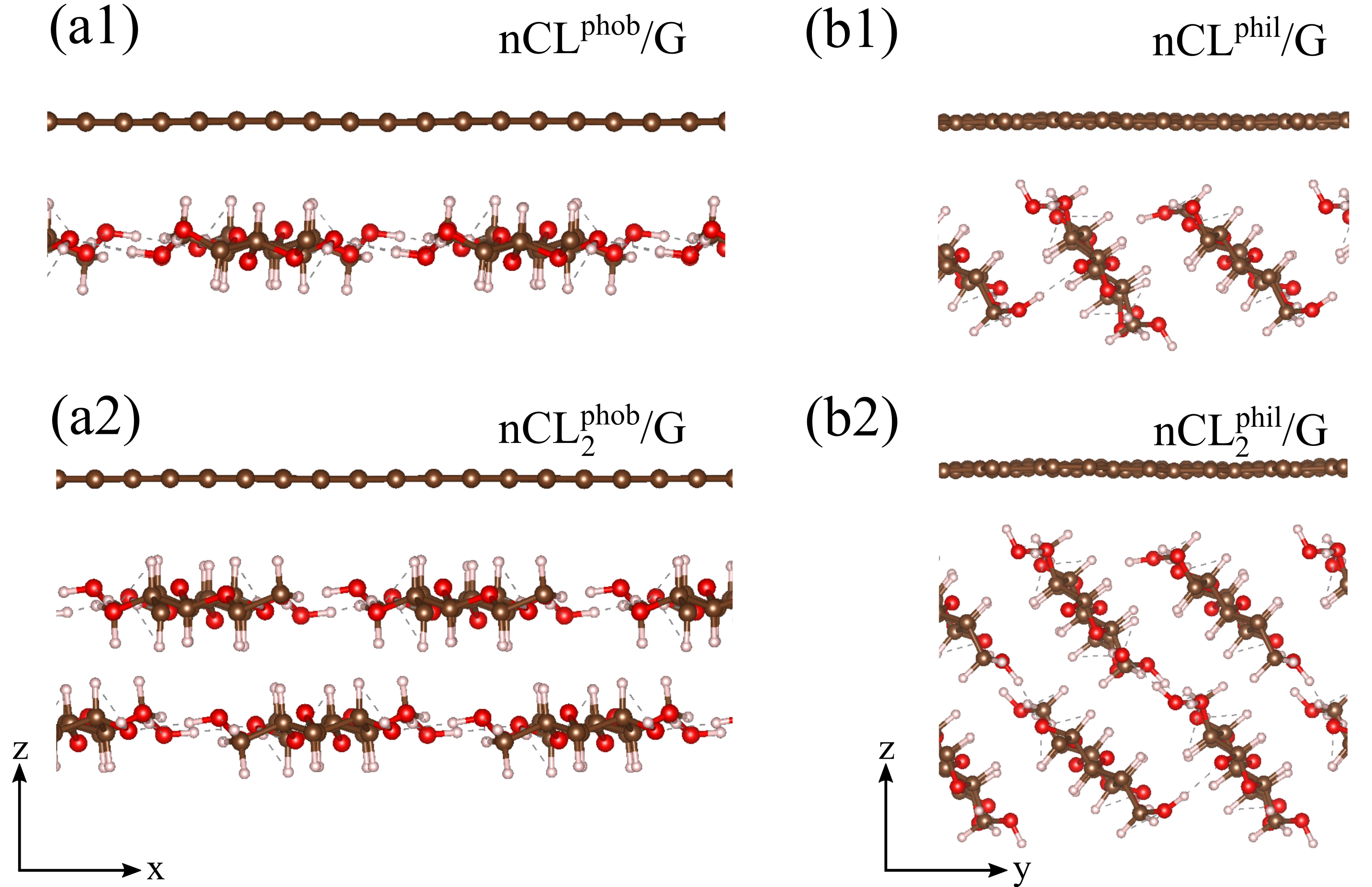}
\caption{\label{models} Structural models of 
graphene interacting with the hydrophobic (a1)-(a2), and   hydrophilic (b1)-(b2) nanocellulose sheet described by a single layer (a1)-(b1), and bilayer (a2)-(b2) of cellulose nanofibrils.}
\end{figure}

\begin{table}
\label{tab:geometry1}
\caption{Binding energies ($E^b$ in meV/\AA$^2$) and the averaged ncl--G vertical 
distance ($h$ in \AA) of the \nclpho/G and \nclphi/G interfaces.}
\begin{ruledtabular}
\begin{tabular}{crcrc}
 & \multicolumn{2}{c}{\nclpho/G} & \multicolumn{2}{c}{\nclphi/G}\\
 \cline{2-3} \cline{4-5}
   vdW     &   $E^b$ &  $h$ &  $E^b$ & $h$  \\
\hline           
  DF       &   12.92  &   2.71  &  11.63  &  2.59 \\ 
  optB86b  &   15.10  &   2.53  &  13.87  &  2.13 \\
  no vdW   &    0.51  &   3.04  &   0.81  &  2.90 \\
DF-solvent &   11.80  &   2.71  &   9.51  &  2.70 \\  
\hline
 & \multicolumn{2}{c}{\nclclpho/G} & 
\multicolumn{2}{c}{\nclclphi/G}\\
 \cline{2-3} \cline{4-5}
   vdW     &   $E^b$ &  $h$ &  $E^b$ & $h$  \\
\hline
    DF     &   13.22   &  2.75  &  12.34 &  2.51 \\
  optB86b  &   16.07   &  2.46  &  13.91 &  2.32 \\
DF-solvent &   12.92   &  2.74  &   9.69 &  2.69 \\  
\end{tabular}
\end{ruledtabular}
\end{table}

We  start our investigation by examining  the energetic stability and  equilibrium geometry of the \ncl/G interface,  and the role played by the long-range vdW forces on the nanocellulose\,-\,graphene binding strength.
 The hydrophobic interface is characterized by the predominance of  CH-$\pi$ bonds, whereas in \nclphi/G the interface interaction is mainly dictated by the  OH-$\pi$ bonds.  The \ncl/G interface binding energy ($E^b$) was calculated by comparing the total energy of the final system ($E[\text{nCL/G}]$) and the sum to the total energies of the isolated components; for example, in Figs.\,\ref{models}(a1) and (b1) a single sheet of  celulose nanofibrils ($E[\text {nCL}]$) and single layer graphene ($E[\text{G}]$),
 \begin{equation}
     E^b = E[{\rm nCL/G}] - E[{\rm nCL}] - E[{\rm G}].\label{b-ene}
 \end{equation}
 For each interface,  i.e. \nclpho/G and \nclphi/G, we have considered three different stacking geometries. We found that $E^b$ and the (averaged) equilibrium vertical distance between the \ncl\, and graphene sheet ($h$) change by less than 0.06\,meV/\AA$^2$ and 0.01\,\AA. 
 
Our results of  $E^b$ and $h$, summarized in Table\,I, reveal  an energetic preference for the  \nclpho/G interface. By using the vdW-DF approach to describe the long-range vdW interactions we found  binding energies of 12.92 and 11.63\,meV/\AA$^2$ for \nclpho/G and \nclphi/G, respectively. Comparing with other layered 2D counterpart systems,  we find that the \nclpho/G binding energy is (i) comparable with that of  boron-nitride/G bilayer ($\sim$12\,meV/\AA$^2$ \cite{fan2011tunable}); (ii) about 13\% smaller compared with the intersheet binding energy of \ncl; \cite{silvestreJPhysChemB2021} and (iii) between 16\% and 40\% higher when compared to graphene oxide (GO) and \nclpho interface, depending on the oxygen concentration \cite{petry2022machine}. In order to verify the reliability of our results, we have also calculated the binding energies (i) by using another vdW functional (optB86b in Table\,I), and  (ii) adding a second layer of cellulose nanofibrils, \nclclpho/G and \nclclphi/G, as depicted in Figs.\,\ref{models}(a2) and (b2), respectively. Our $E^b$ results, Table\,I, confirm the energetic preference for the \nclpho/G interface.

The energetic preference of the \nclpho/G interface is in agreement with recent theoretical findings based on molecular dynamic simulations \cite{alqusBiomacro2015, mianehrow2022interface}. Both studies indicate that the presence of trapped water molecules at the \ncl/G interface reduces the binding energy; moreover in Ref.\,\onlinecite{alqusBiomacro2015} the authors verified the exclusion of the water molecules from the \nclpho/G interface. Indeed, by using the implicit solvation model \cite{mathew2014implicit, mathew2019implicit}  we find a reduction of the interface binding energy [Eq.\,\ref{b-ene}], $E^b$\,=\,12.92\,$\rightarrow$\,11.80\,meV/\AA$^2$ in \nclpho/G, and 11.63\,$\rightarrow$\,9.51\,meV/\AA$^2$ in the \nclphi/G interface, while the equilibrium geometries of the \ncl/G interfaces are nearly the same as those obtained previously with no solvent effects. It is worth noting that the larger binding energy reduction in the latter is due to the hydrophilic nature of the interface, which is in agreement with our results of solvation energies ($E^s$)\footnote{$E^s$ is defined as the diference between the vacuum total energy and solvent total energy.}, namely 2.87 and 10.77\,meV/\AA$^2$ in \nclpho/G and \nclphi/G, respectively, and 4.21 and 11.19 in the nanocellulose bilayer systems, \nclclpho/G and \nclclphi/G.

Long range vdW dispersion interaction plays an important role in the interchain and intersheet binding energy between the cellulose fibrils and \ncl\, sheets, respectively \cite{liJPhysChemC2011, silvestreJPhysChemB2021}. To quantify the role played by the vdW interaction on the \ncl\,-\,G binding energy, we have calculated $E^b$ by turning off the vdW contribution. In this case, the binding energy of the \nclpho/G (\nclphi/G) interface  reduces to 0.51 (0.81)\,meV/\AA$^2$, and the vertical distance $h$ increases to 3.04 (2.90)\,\AA. Thus, similarly to the inter-sheet binding energy in pristine \ncl,\cite{silvestreJPhysChemB2021} we can deduce that the non-covalent (vdW) interactions rules the  formation of \ncl/G interfaces.

\begin{figure}
\includegraphics[width=\columnwidth]{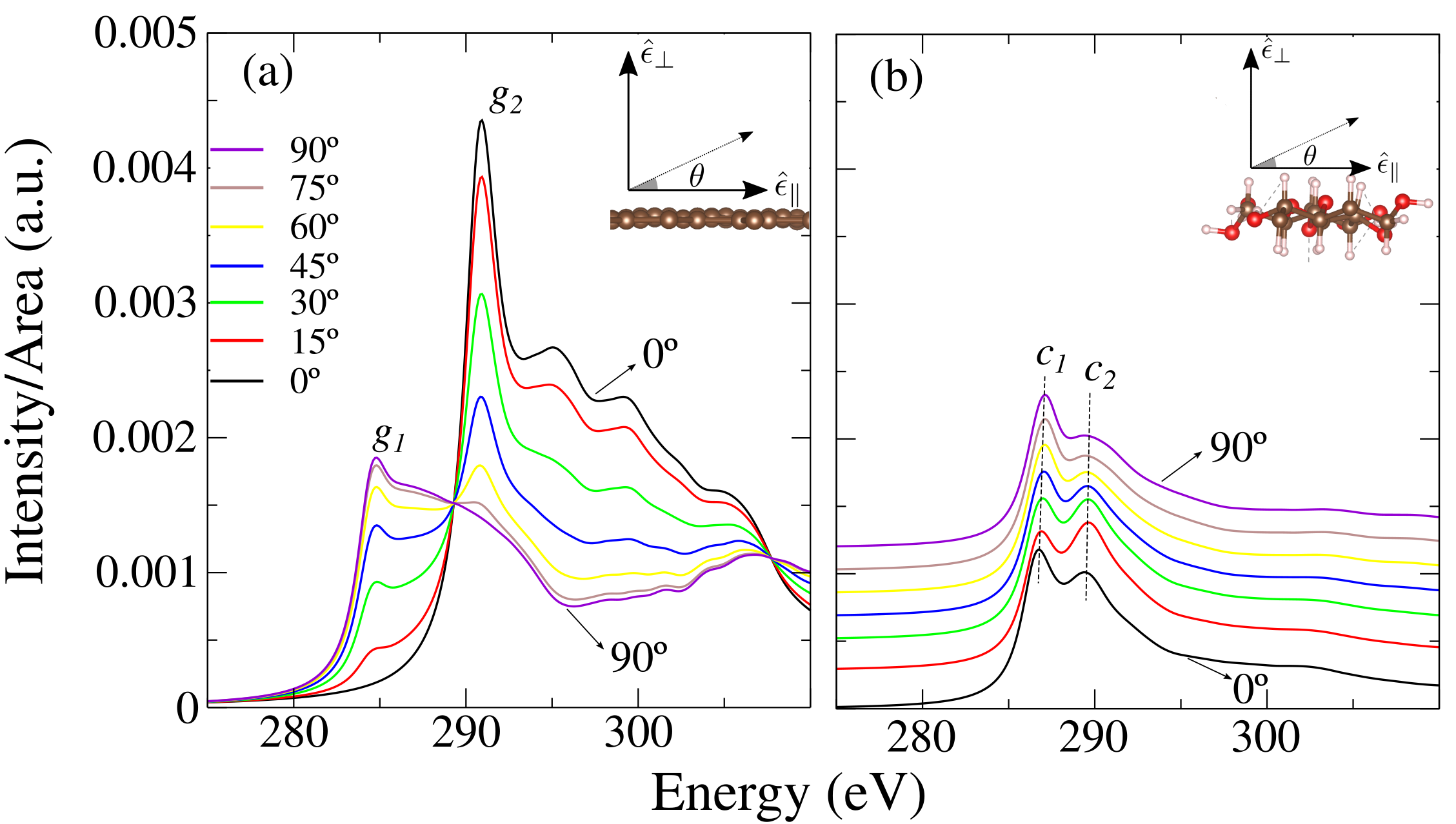}
\caption{\label{xas-full-all} Carbon K-edge simulated XANES spectra of pristine graphene\,(a), and single layer cellulose fibrils (b) as a function of the radiation polarization angle ($\theta$).} 
\end{figure}
  
\subsection{Structural Characterization, XANES}  

Core-level spectroscopy is a powerful tool to provide the structural characterization of materials in an atomic scale based on the local electronic properties of the probed element. Currently, the combination of experimental X-ray absorption near edge structure (XANES) data and first-principles simulations has proven to be a highly successful strategy to understand the atomic structure of novel materials \cite{prendergast2006x, xas2, de2020simulations, inayeh2021self}. In this subsection, we present  XANES simulation results of the Carbon K-edge absorption spectra of \ncl/G interfaces in order to find spectroscopic finger prints of the atomic structures of the \nclpho/G and \nclphi/G interfaces.

Let us start with the XANES of the pristine isolated systems, graphene and single layer \ncl.  In Fig.\,\ref{xas-full-all}(a), we present the absorption spectra of graphene as a function of the orientation ($\theta$) of the radiation polarization vector ($\hat\epsilon$), where we can identify  the C-1$s$\,$\rightarrow$\,$\pi^{\ast}$ [$\rightarrow$\,$\sigma^{\ast}$] transition for $\hat\epsilon=\hat\epsilon_\perp$ ($\theta$\,=\,90$^\circ$) [$\hat\epsilon=\hat\epsilon_\parallel$ ($\theta$\,=\,0$^\circ$)]. The energy positions of these absorption peaks, around 285 and  292\,eV, respectively, indicated as $g_1$ and $g_2$ in Fig.\,\ref{xas-full-all}(a), and the dependence of their intensities with the orientation of the polarization vector are in good agreement with the previous experimental findings \cite{schiros2012connecting, lippitz2013plasma}.

\begin{figure}
\includegraphics[width=\columnwidth]{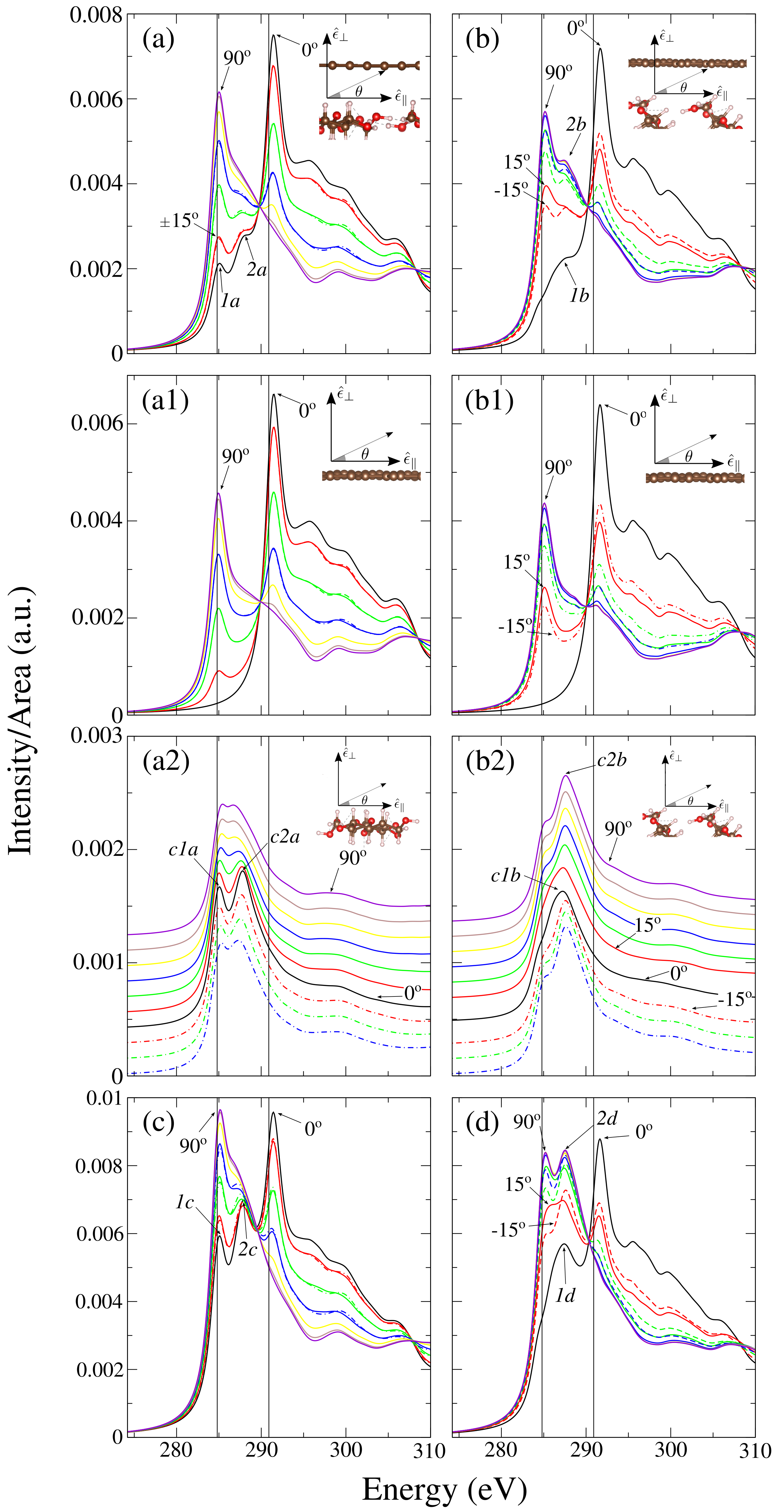}
\caption{\label{xas-full-all2} XANES spectra of \nclpho/G (a), and \nclphi/G (b) interfaces. XANES spectra of hypothetical graphene [(a1)-(b1)] and single layer \ncl\, sheet [(a2)-(b2)] constrained to the  equilibrium geometry of the respective final system, \nclpho/G and \nclphi/G, as indicated in the insets. XANES spectra of \nclclclpho/G (c), and \nclclclphi/G (d).} 
\end{figure}

Experimental results of XANES spectra of cellulose \cite{cody2000probing, karunakaran2015introduction} indicate the presence of two absorption peaks, at  289.3 and 290.7\,eV, both attributed to the  C-1$s$\,$\rightarrow$\,$\pi^{\ast}$ transition, associated to the C--OH and C--H bonds, respectively. Those bonding structures are present in the single layer \ncl, and, indeed, the respective absorption spectra  are captured in our simulations. In Fig.\,\ref{xas-full-all}(b), we show our results for a single-layer \ncl\, sheet which are characterized by (i) two absorption peaks, denoted $c_1$ and $c_2$, lying at about 287 and 290\,eV, and (ii) reduced angular dependency with the direction of the polarization vector compared to the one of graphene [Fig.\,\ref{xas-full-all}(a)]. As shown in Fig.\,\ref{xas-full-all}(b), the energy positions of $c_1$ and $c_2$ changes by 0.36 and 0.21\,eV, respectively, for $\theta=0\rightarrow 90^\circ$.\footnote{The simulations of the XANES spectra of \ncl\, were performed by taken into account the polar angle [$\theta$, presented in Figs.\ref{xas-full-all} and \ref{xas-full-all2}], and two azimuthal ($\phi$) angles, one for polarization vectors parallel and another perpendicular  to the cellulose fibrils.} The absorption spectra in the \ncl\, bulk phase  (not shown) have a similar characteristic, with $c_1$ and $c_2$ shifted by $\sim$\,+2 eV, for example $c1$: 287\,$\rightarrow$\,289 eV. 

The XANES spectra of \nclpho/G and \nclphi/G interfaces are shown in Figs.\,\ref{xas-full-all2}(a) and (b). In both spectra we identify the C-$1s$\,$\rightarrow$\,$\pi^\ast$ and $\rightarrow$\,$\sigma^\ast$ absorption features from graphene for $\theta=90^\circ$ and $0^\circ$, respectively. Between these two transitions energies, we can identify the following differences on the absorption features in \nclpho/G and \nclphi/G attributed to the \ncl\, absorption spectra.  (i) The presence of absorption peaks {\it 1a} and {\it 2a} [Fig.\,\ref{xas-full-all2}(a)] for $\theta=0^\circ$,  while in the \nclphi/G we find one absorption feature, {\it 1b} in Fig.\,\ref{xas-full-all2}(b); (ii)  the absorption peak  {\it 2b} at 287.5\,eV is clearly visible in the \nclphi/G interface for $\theta=90^\circ$, but it is attenuated in \nclpho/G; and (iii)  due to the tilted geometry of cellulose fibrils with respect to the graphene layer, the absorption spectra  in \nclphi/G are no longer symmetric for positive and negative values of radiation polarization angles. For instance, the absorption features for $\theta=+15^\circ$ and $-15^\circ$ (dot-dashed lines) present different intensities, as seen  in Fig.\,\ref{xas-full-all2}(b). 

Because the formation of \ncl/G heterostruture is ruled by the vdW interactions, with no covalent bonds between the \ncl\, sheet and the graphene layer, the \ncl/G interfaces' absorption spectra are primarily determined by the superposition of the ones of isolated components. Indeed, the absorption spectra can be better understood by XANES computations of hypothetical structures, namely, an isolated single-layer graphene, and \ncl\, sheet (both) constrained to the respective \ncl/G interface equilibrium geometry. The XANES simulations of these constrained structures  reveal that  the features {\it 1a} and {\it 2a} in \nclpho/G emerge from the superposition of the edge transitions in graphene with the absorption peaks {\it c1a} and {\it c2a} of the \ncl, Figs.\,\ref{xas-full-all2}(a1) and (a2). Similarly, the absorption peak {\it 1b}  in \nclphi/G  results from the superposition of graphene edge absorption structure with the {\it c1b} peak of the tilted layer of cellulose fibrils, Figs.\,\ref{xas-full-all2}(b1) and (b2), and  for $\theta=90^\circ$ the absorption feature {\it 2b} is composed by the superposition of graphene near-edge structure with the absorption peak {\it c2b} of the tilted \ncl. 

The absorption spectra of the \ncl/G systems, Figs.\,\ref{xas-full-all2}(a) and (b),  were calculated by considering graphene on a single layer of cellulose sheet. However, when the number of \ncl\, sheets grows, it is worthwhile to examine the changes on the XANES spectra. Our results reveal that  as the number of \ncl\, layers increases, the absorption characteristics from the C\,-\,H and C\,-\,OH bonds grow more pronounced between 285 and 292\,eV. In Fig.\,\ref{xas-full-all2}(c) and (d) we present the absorption spectra   for three layer of cellulose fibrils, \nclclclpho/G and \nclclclphi/G. In the former interface  the absorption peaks {\it 1a} and {\it 2a} become more apparent [labelled as {\it 1c} and {\it 2c} in Fig.\,\ref{xas-full-all2}(c)], similarly the features {\it 1b} and {\it 2b} of \nclphi/G   become more intense in \nclclclphi/G, indicated as {\it 1d} and {\it 2d} in Fig.\,\ref{xas-full-all2}(d), and thus reinforcing the differences in the  XANES signatures of the hydrophobic and hydrophilic \ncl/G interfaces.

\begin{figure}
\includegraphics[width=\columnwidth]{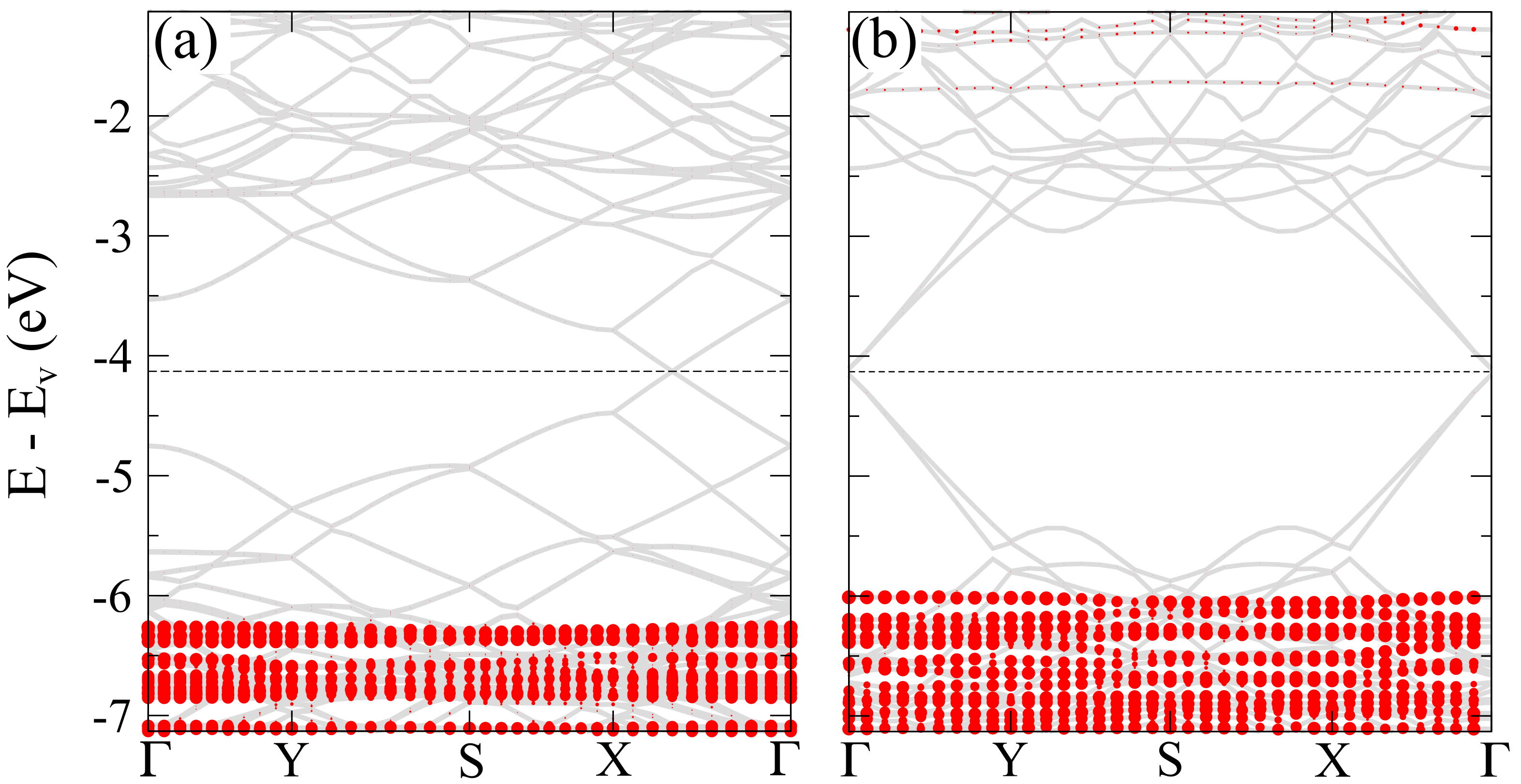}
\caption{\label{bandas} Electronic band structure of \nclpho/G (a) and \nclphi/G (b). The zero energy was set at the vacuum level, the Fermi level is indicated by the dashed black lines, and red circles indicates cellulose contribution in band.} 
\end{figure}
 
\subsection{Electronic properties} 

The electronic properties of \ncl/G are characterized by an insulator/semi-metal interface with the   linear energy bands of graphene lying within the bandgap of \ncl. In Figs.\,\ref{bandas}(a) and (b), we present the orbital projected electronic band structures of \nclpho/G and \nclphi/G, where  we  find (i) the graphene's Dirac-point (DP)  at about 2\,eV above the  valence band maximum (VBM) of the \ncl\, layer; and (ii) the emergence of an energy gap of $\sim$0.04\,eV at the DP. There is a small amount of charge accumulation at the \ncl/G interface. Based on the Bader analysis \cite{bader}, we found a net charge transfer  ($\Delta\rho$)  of $0.23\,(0.28)\times 10^{13}$\,$e$/cm$^2$ from graphene to the \nclpho/G, (\nclphi/G) interface, followed by a reduction of the graphene work-function  ($\Phi$)\footnote{The work function, $\Phi$, is defined as the energy position  of the Fermi level with respect to the vacuum level, $\Phi=E_{\rm F}-E_{\rm vac}$, where $E_{\rm vac}$  is obtained from the electrostatic potential calculation in a vacuum region far away from the system. Here we set $E_{\rm vac}=0$.} by  0.2\,eV compared with that of free-stading graphene sheet, 4.33\,$\rightarrow$\,4.13\,eV. Similar results were obtained in the \nclclpho/G and \nclclphi/G interfaces. 

\begin{figure}
\includegraphics[width=\columnwidth]{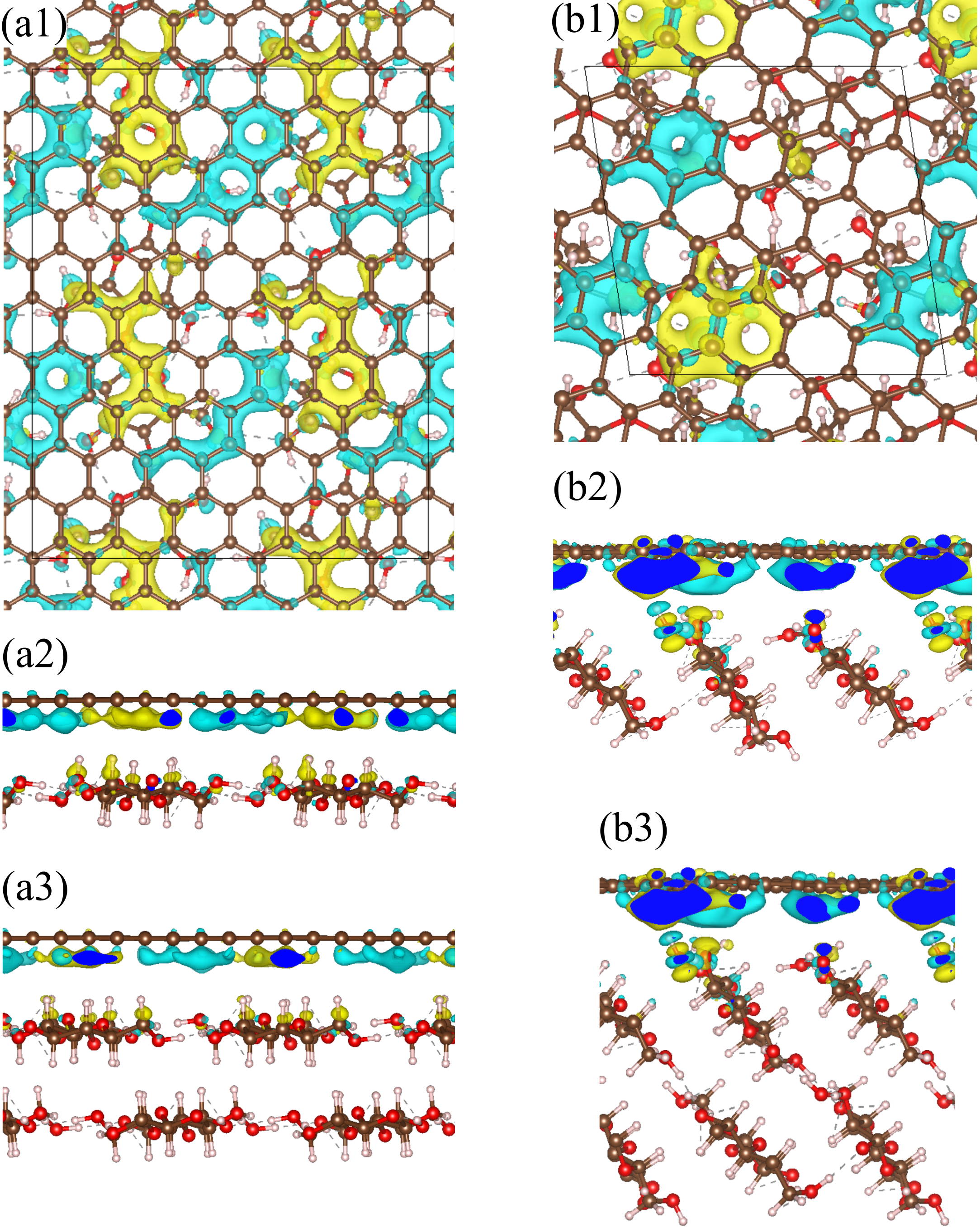}
\caption{\label{deltarho} Net charge tranfers, $\Delta\rho$  in \nclpho/G (a1), \nclphi/G (b1), \nclclpho/G (a2), and \nclclphi/G (b2). Isosurfaces of 0.4 me/\AA$^{3}$, and postive (negative) values of $\Delta\rho$ are indicated in blue (yellow).}
\end{figure}

In Figs.\,\ref{deltarho}(a) an (b) we present a map of the charge accumulations in the \nclpho/G and \nclphi/G interfaces, where the following characteristics  are noteworthy: (i) the inhomogeneous (net) charge distribution on the graphene sheets [Figs.\,\ref{deltarho}(a1) and (b1)], which can be attributed to the differences in the orbital hopping  between the \ncl\, surface and the  graphene's $\pi$ orbitals; and (ii), as depicted in Figs.\,\ref{deltarho}(a2) and (b2), those charge transfers occurs primarily at the \ncl/G interface, since $\Delta\rho\approx 0$ in the subsurface \ncl\,layers, Figs.\,\ref{deltarho}(a3) and (b3). {Such a net charge localization suggest the emergence of localized electronic transmission channels at the \ncl/G interface. On the other hand, it} is worth mentioning that such an inhomogeneous net charge distribution [(i)], giving rise to electron- and hole-rich regions in graphene, will play a deleterious role on the electronic transport  properties throughout the {\ncl/G} layers. 

The understanding of the electronic properties of the \ncl/G interface in the presence of external agents is an important issue for the development of electronic devices/sensors based on a combination of cellulosic materials and graphene. Thus, in the sequence, we will focus on the effects of  external electric field (EEF) and mechanical compressive strain in the \ncl/G interface.

\begin{figure}
\centering
\includegraphics[width=\columnwidth]{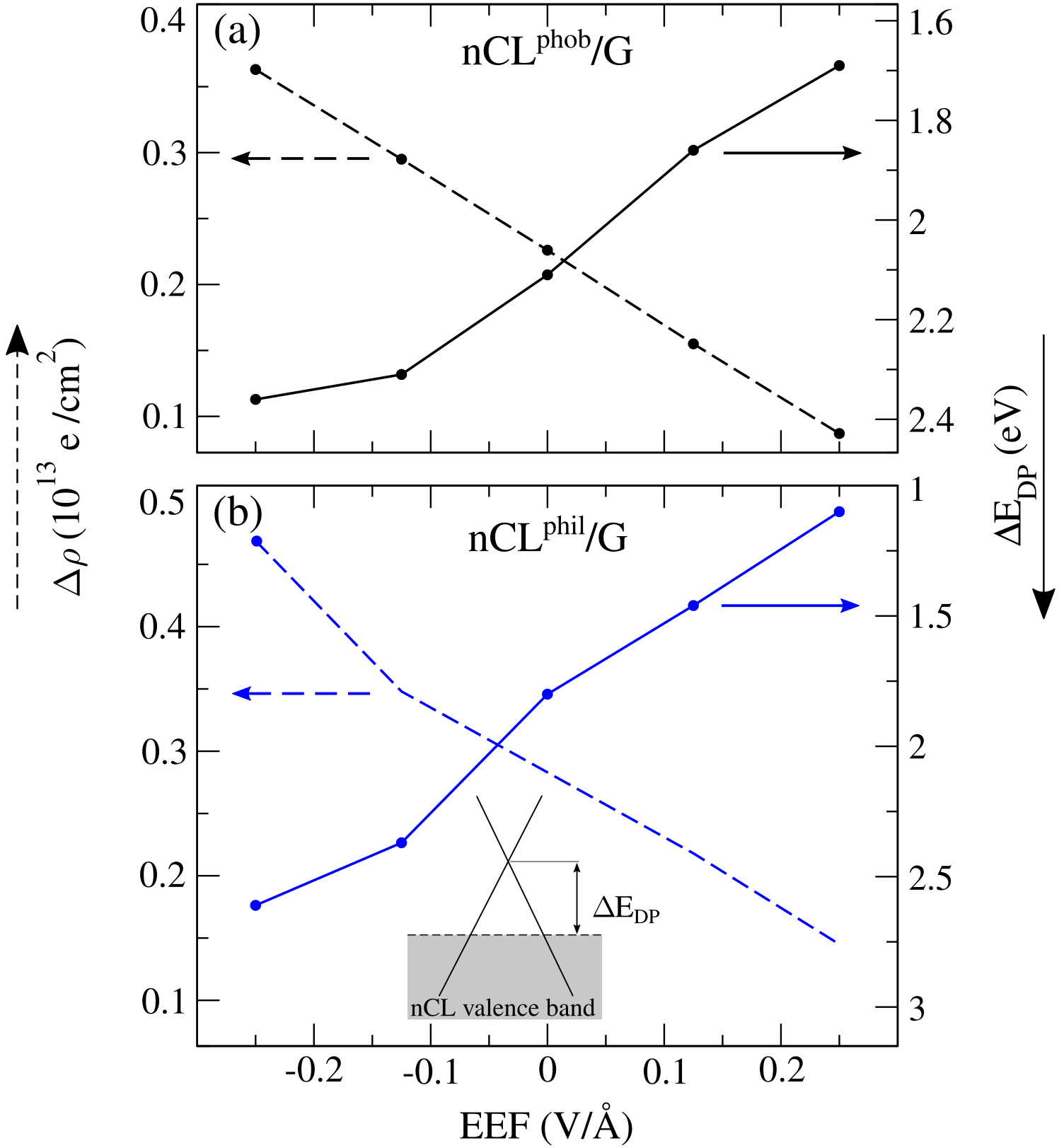}
\caption{\label{rho-DP} Energy position of the DP with respect to the VBM ($\Delta E_{\rm DP}$ - filled lines), and the net charge transfer from graphene to the \ncl\, ($\Delta\rho$ in 10$^{13}\,e$/cm$^2$ - dashed lines) as a function of the external electric field (EEF) for the  \nclpho/G (a), and \nclphi/B (b) interfaces.} 
\end{figure}

Let us start with the effect of EEF.\footnote{For each value of electric field, the atomic positions of the \nclpho/G system were fully relaxed.}  As shown in Fig.\,\ref{rho-DP}, the energy position of the Dirac point with respect to the VBM of the \ncl\, ($\Delta E_{\rm DP}$) increases from 1.6 (1.1) to 2.3 (2.6)\,eV in \nclpho/G (\nclphi/G), for EEF of $-0.25$ and $+0.25$\,V/\AA, respectively. Concomitantly, the G\,$\rightarrow$\,\ncl/G\, net charge transfer, $\Delta\rho$, varies from 0.08 (0.14) to 0.36 (0.47)\,$\times 10^{13}\,e/\text{cm}^2$. Such tuning of $\Delta E_{\rm DP}$ and $\Delta\rho$ is nearly linear with respect to the EEF. In particular, for $\Delta E_{\rm DP}$ as a function of the EEF we find the following rates,  $-1.43$\,eV/(V/\AA) and $-3.14$\,eV/(V/\AA) in \nclpho/G and \nclphi/G, respectively. Assuming that such a linear relationship is preserved for larger values of EEF, we can infer that in \nclphi/G  the DP becomes resonant with the \ncl\, valence band for EEF\,$>$\,0.6\,V/\AA, and thus suppressing the G\,$\rightarrow$\,\ncl/G\, charge transfer. 

\begin{figure}
\includegraphics[width=\columnwidth]{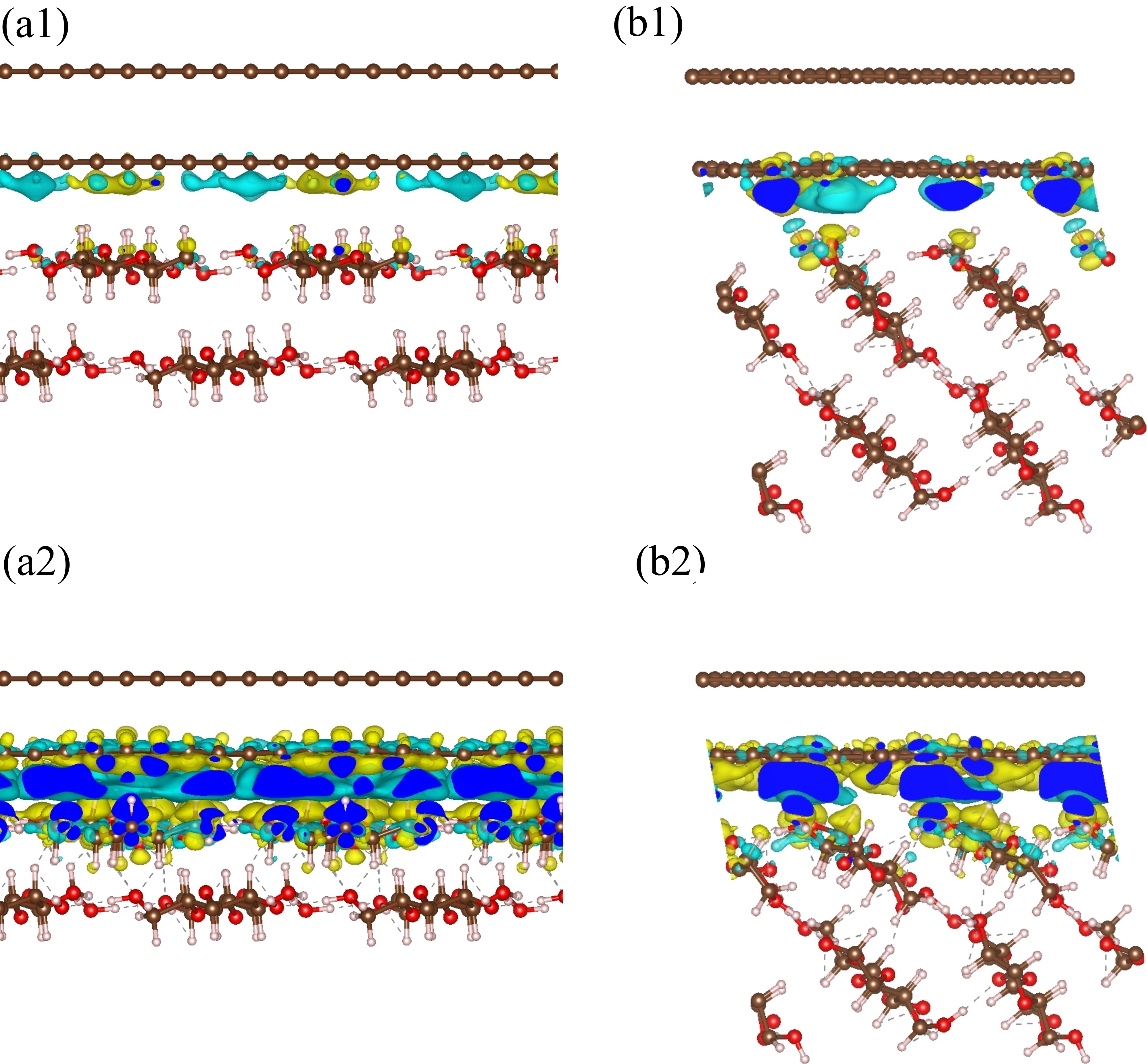}
\caption{\label{posits-compress} Side view of differential charge densities of \nclclpho/GBL (a) to 9\% z-compressed \nclclpho/GBL (b), and \nclclphi/GBL (c) to 9\% z-compressed \nclclphi/GBL (d). Isosurfaces of 0.4 me/\AA$^{3}$}
\end{figure}

\begin{figure}[h]
\centering
\includegraphics[width=\columnwidth]{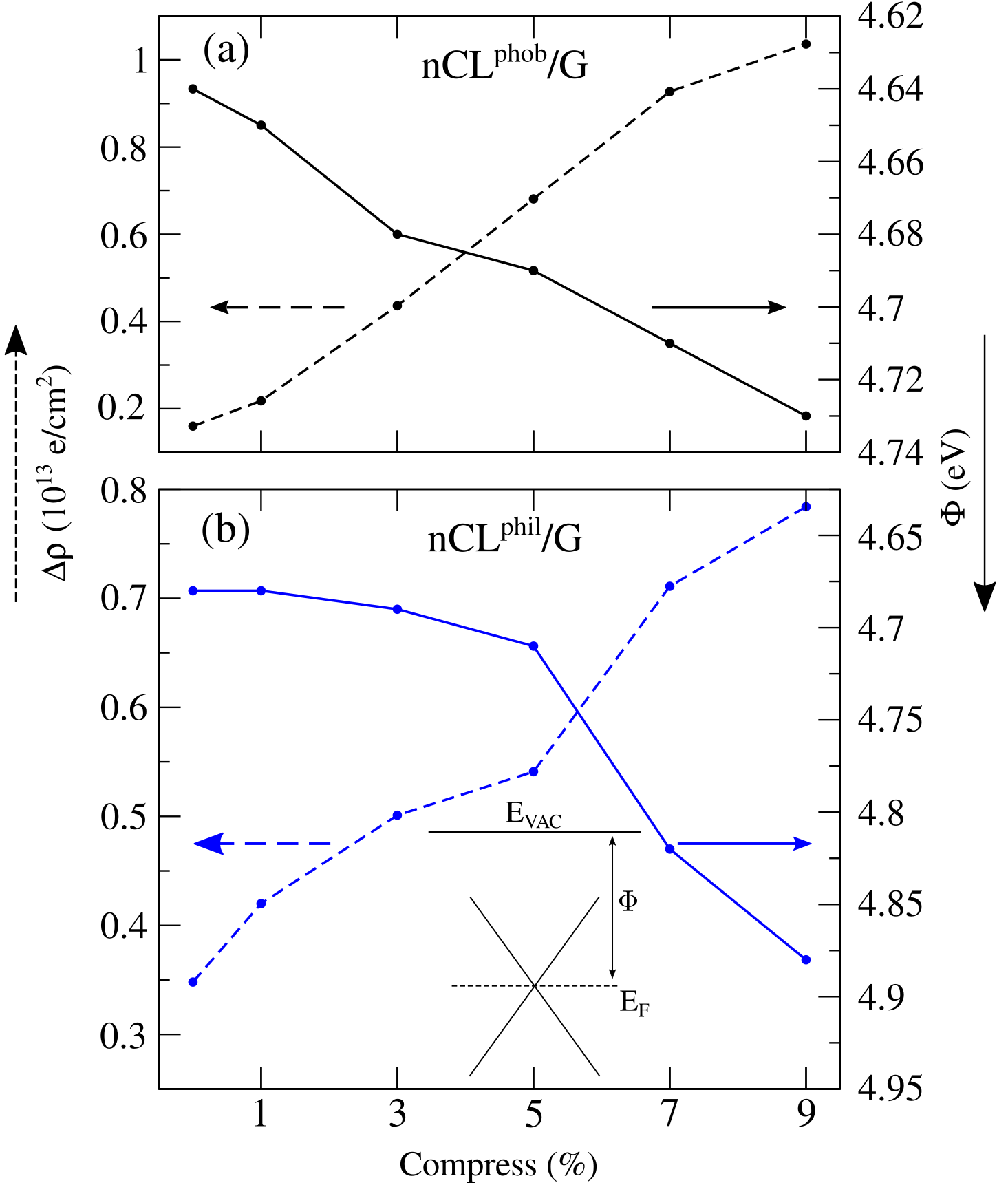}
\caption{\label{WF} Net charge tranfer, $\Delta\rho$, and the work function $\Phi$ upon compression of \nclclpho/GBL (a) and \nclclphi/GBL (b). E$_\text{vac}$ represents the vacuum level.} 
\end{figure}

Further control of the electronic properties of 2D systems can be achieved through mechanical strain, ``straintronics" \cite{si2016strain, miao2021straintronics}. Indeed, such an approach has been used to control the electronic doping level in bilayer graphene and boron-nitride/graphene vdW  heterostructures \cite{yankowitz2016pressure, vincent2018probing, forestier2020strain}. {In \ncl/G, the compressive strain will promote the strengthening the CH-$\pi$ (OH-$\pi$) orbital overlap at the \nclpho/G (\nclphi/G) interface.} Here, we investigate the net charge transfer, $\Delta\rho$, and the work function ($\Phi$) in \ncl/G, as a function of  compressive strain.  The strain in the \nclpho/G and \nclphi/G interfaces was applied by considering bilayers of \ncl\, and graphene,  \nclclpho/GBL [Fig.\,\ref{posits-compress}(a)] and  \nclclphi/GBL [Fig.\,\ref{posits-compress}(b)]. In Figs.\,\ref{posits-compress}(a1) and (a2) we present the spatial distribution of $\Delta\rho$  of pristine \nclclpho/GBL  and compressed by 9\,\%, respectively; similarly in Figs.\,\ref{posits-compress}(b1) and (b2), we present $\Delta\rho$ for the hydrofilic interface, \nclclphi/GBL. Interestingly, these  $\Delta\rho$ maps reveal that the charge transfers are localized at the \nclcl/GBL interface region, and as shown in Figs.\,\ref{WF}, the net charge transfer from G to the \ncl\, increases up to $\sim$\,1.0 ($\sim$\,0.8)\,$\times 10^{13}\,e/\text{cm}^2$ in \nclclpho/GBL  (\nclclphi/GBL) for a compressive strain of about 9\,\%, which corresponds to an external pressure of 3.73 (3.15)\,GPa. Concomitantly with the G\,$\rightarrow$\,\ncl/G\, net charge transfer, we find slight increase of the work functions, namely from 4.64 (4.68) to 4.73 (4.88)\,eV in \nclclphi/GBL (\nclclpho/GBL) [Fig.\,\ref{WF}]. {Finally,  we found that we can reach a $p$-type doping of  about 1.1 (1.3)\,$\times 10^{13}\,e/\text{cm}^2$ of graphene upon an EEF of $-0.25$\,V/\AA\, in the 9\% compressed \nclpho/G\, (\nclphi/G) interfaces. Thus,  suggesting that suitable combinations of these external agents can be exploited to further modify the electrical characteristics of the \ncl/G interface.}

\section{Summary and Conclusions}

We have performed a theoretical investigation of the energetic, structural, and electronic properties of nanocellulose/graphene (\ncl/G) interface, where we have addressed the hydrophobic (\nclpho/G) and hydrophilic (\nclphi/G) interfaces. We find that the binding energy of \ncl/G is primarily ruled by the vdW interactions, being comparable with that of boron-nitride/graphene. The structural fingerprints of \nclpho/G and \nclphi/G interfaces were identified through a detailed study of the Carbon K-edge absorption (XANES) spectra. The electronic structure of \ncl/G is characterized by linear energy bands of graphene lying within the bandgap of \ncl, with the Dirac point at about 2\,eV above the valence band maximum of the \ncl\, sheet, $\Delta E_\text{DP}\approx 2\,\text{eV}$, and a net charge accumulation, $\Delta\rho$ of $\sim{0.2}\times 10^{13} e/\text{cm}^2$,  localized at the \ncl/G interface. External electric fields (EEFs) and mechanical strain were used to investigate the tunability of these quantities, where we found that $\Delta E_{\text{DP}}$  varies from 1.6 (1.1) to 2.3 (2.6)\,eV in \nclpho/G (\nclphi/G), for EEF of $-0.25$ and $+0.25$\,V/\AA, respectively; whereas there is an increase of G$\rightarrow$\,\ncl/G $\Delta\rho$ up to $1\times 10^{13} e/{\text{cm}}^2$  upon an external pressure of 3.73\,GPa. We believe that {our} findings provide not only the energetic and atomic scale structural understanding of the \ncl/G interface, but also a set of important  information of the electronic properties to the development of electronic devices based on the combination of nanocellulose and graphene.

\begin{acknowledgments}

The authors acknowledge   financial   support   from   the Brazilian  agencies  CNPq, FAPEMIG, and FAPESP (grant 16/04514-7 and 17/02317-2), and the  Laborat\'orio Nacional de Computa\c{c}\~ao Cient\'{\i}fica (LNCC-SCAFMat2), Centro Nacional de Processamento de Alto Desempenho (CENAPAD-SP) for computer time.

\end{acknowledgments}

\bibliography{bib}

\begin{thebibliography}{63}%
\makeatletter
\providecommand \@ifxundefined [1]{%
 \@ifx{#1\undefined}
}%
\providecommand \@ifnum [1]{%
 \ifnum #1\expandafter \@firstoftwo
 \else \expandafter \@secondoftwo
 \fi
}%
\providecommand \@ifx [1]{%
 \ifx #1\expandafter \@firstoftwo
 \else \expandafter \@secondoftwo
 \fi
}%
\providecommand \natexlab [1]{#1}%
\providecommand \enquote  [1]{``#1''}%
\providecommand \bibnamefont  [1]{#1}%
\providecommand \bibfnamefont [1]{#1}%
\providecommand \citenamefont [1]{#1}%
\providecommand \href@noop [0]{\@secondoftwo}%
\providecommand \href [0]{\begingroup \@sanitize@url \@href}%
\providecommand \@href[1]{\@@startlink{#1}\@@href}%
\providecommand \@@href[1]{\endgroup#1\@@endlink}%
\providecommand \@sanitize@url [0]{\catcode `\\12\catcode `\$12\catcode
  `\&12\catcode `\#12\catcode `\^12\catcode `\_12\catcode `\%12\relax}%
\providecommand \@@startlink[1]{}%
\providecommand \@@endlink[0]{}%
\providecommand \url  [0]{\begingroup\@sanitize@url \@url }%
\providecommand \@url [1]{\endgroup\@href {#1}{\urlprefix }}%
\providecommand \urlprefix  [0]{URL }%
\providecommand \Eprint [0]{\href }%
\providecommand \doibase [0]{http://dx.doi.org/}%
\providecommand \selectlanguage [0]{\@gobble}%
\providecommand \bibinfo  [0]{\@secondoftwo}%
\providecommand \bibfield  [0]{\@secondoftwo}%
\providecommand \translation [1]{[#1]}%
\providecommand \BibitemOpen [0]{}%
\providecommand \bibitemStop [0]{}%
\providecommand \bibitemNoStop [0]{.\EOS\space}%
\providecommand \EOS [0]{\spacefactor3000\relax}%
\providecommand \BibitemShut  [1]{\csname bibitem#1\endcsname}%
\let\auto@bib@innerbib\@empty
\bibitem [{\citenamefont {Irimia-Vladu}(2014)}]{irimia2014green}%
  \BibitemOpen
  \bibfield  {author} {\bibinfo {author} {\bibfnamefont {Mihai}\ \bibnamefont
  {Irimia-Vladu}},\ }\bibfield  {title} {\enquote {\bibinfo {title}
  {{“Green” electronics: biodegradable and biocompatible materials and
  devices for sustainable future}},}\ }\href {\doibase 10.1039/C3CS60235D}
  {\bibfield  {journal} {\bibinfo  {journal} {Chem. Soc. Rev.}\ }\textbf
  {\bibinfo {volume} {43}},\ \bibinfo {pages} {588--610} (\bibinfo {year}
  {2014})}\BibitemShut {NoStop}%
\bibitem [{\citenamefont {Zhu}\ \emph {et~al.}(2016)\citenamefont {Zhu},
  \citenamefont {Luo}, \citenamefont {Ciesielski}, \citenamefont {Fang},
  \citenamefont {Zhu}, \citenamefont {Henriksson}, \citenamefont {Himmel},\
  and\ \citenamefont {Hu}}]{zhu2016wood}%
  \BibitemOpen
  \bibfield  {author} {\bibinfo {author} {\bibfnamefont {Hongli}\ \bibnamefont
  {Zhu}}, \bibinfo {author} {\bibfnamefont {Wei}\ \bibnamefont {Luo}}, \bibinfo
  {author} {\bibfnamefont {Peter~N.}\ \bibnamefont {Ciesielski}}, \bibinfo
  {author} {\bibfnamefont {Zhiqiang}\ \bibnamefont {Fang}}, \bibinfo {author}
  {\bibfnamefont {J.~Y.}\ \bibnamefont {Zhu}}, \bibinfo {author} {\bibfnamefont
  {Gunnar}\ \bibnamefont {Henriksson}}, \bibinfo {author} {\bibfnamefont
  {Michael~E.}\ \bibnamefont {Himmel}}, \ and\ \bibinfo {author} {\bibfnamefont
  {Liangbing}\ \bibnamefont {Hu}},\ }\bibfield  {title} {\enquote {\bibinfo
  {title} {{Wood-Derived Materials for Green Electronics, Biological Devices,
  and Energy Applications}},}\ }\href {\doibase 10.1021/acs.chemrev.6b00225}
  {\bibfield  {journal} {\bibinfo  {journal} {Chemical Reviews}\ }\textbf
  {\bibinfo {volume} {116}},\ \bibinfo {pages} {9305--9374} (\bibinfo {year}
  {2016})}\BibitemShut {NoStop}%
\bibitem [{\citenamefont {Yao}\ \emph {et~al.}(2017)\citenamefont {Yao},
  \citenamefont {Zhang}, \citenamefont {Kou}, \citenamefont {Song},
  \citenamefont {Liu},\ and\ \citenamefont {Li}}]{yao2017paper}%
  \BibitemOpen
  \bibfield  {author} {\bibinfo {author} {\bibfnamefont {Bin}\ \bibnamefont
  {Yao}}, \bibinfo {author} {\bibfnamefont {Jing}\ \bibnamefont {Zhang}},
  \bibinfo {author} {\bibfnamefont {Tianyi}\ \bibnamefont {Kou}}, \bibinfo
  {author} {\bibfnamefont {Yu}~\bibnamefont {Song}}, \bibinfo {author}
  {\bibfnamefont {Tianyu}\ \bibnamefont {Liu}}, \ and\ \bibinfo {author}
  {\bibfnamefont {Yat}\ \bibnamefont {Li}},\ }\bibfield  {title} {\enquote
  {\bibinfo {title} {{Paper-Based Electrodes for Flexible Energy Storage
  Devices}},}\ }\href {\doibase https://doi.org/10.1002/advs.201700107}
  {\bibfield  {journal} {\bibinfo  {journal} {Advanced Science}\ }\textbf
  {\bibinfo {volume} {4}},\ \bibinfo {pages} {1700107} (\bibinfo {year}
  {2017})}\BibitemShut {NoStop}%
\bibitem [{\citenamefont {Zhao}\ \emph {et~al.}(2021)\citenamefont {Zhao},
  \citenamefont {Zhu}, \citenamefont {Cheng}, \citenamefont {Chen},
  \citenamefont {Wu},\ and\ \citenamefont {Yu}}]{zhao2021cellulose}%
  \BibitemOpen
  \bibfield  {author} {\bibinfo {author} {\bibfnamefont {Dawei}\ \bibnamefont
  {Zhao}}, \bibinfo {author} {\bibfnamefont {Ying}\ \bibnamefont {Zhu}},
  \bibinfo {author} {\bibfnamefont {Wanke}\ \bibnamefont {Cheng}}, \bibinfo
  {author} {\bibfnamefont {Wenshuai}\ \bibnamefont {Chen}}, \bibinfo {author}
  {\bibfnamefont {Yiqiang}\ \bibnamefont {Wu}}, \ and\ \bibinfo {author}
  {\bibfnamefont {Haipeng}\ \bibnamefont {Yu}},\ }\bibfield  {title} {\enquote
  {\bibinfo {title} {{Cellulose-Based Flexible Functional Materials for
  Emerging Intelligent Electronics}},}\ }\href {\doibase
  https://doi.org/10.1002/adma.202000619} {\bibfield  {journal} {\bibinfo
  {journal} {Advanced Materials}\ }\textbf {\bibinfo {volume} {33}},\ \bibinfo
  {pages} {2000619} (\bibinfo {year} {2021})}\BibitemShut {NoStop}%
\bibitem [{\citenamefont {Conti}\ \emph {et~al.}(2020)\citenamefont {Conti},
  \citenamefont {Pimpolari}, \citenamefont {Calabrese}, \citenamefont
  {Worsley}, \citenamefont {Majee}, \citenamefont {Polyushkin}, \citenamefont
  {Paur}, \citenamefont {Pace}, \citenamefont {Keum}, \citenamefont {Fabbri},
  \citenamefont {Iannaccone}, \citenamefont {Macucci}, \citenamefont {Coletti},
  \citenamefont {Mueller}, \citenamefont {Casiraghi},\ and\ \citenamefont
  {Fiori}}]{conti2020low}%
  \BibitemOpen
  \bibfield  {author} {\bibinfo {author} {\bibfnamefont {Silvia}\ \bibnamefont
  {Conti}}, \bibinfo {author} {\bibfnamefont {Lorenzo}\ \bibnamefont
  {Pimpolari}}, \bibinfo {author} {\bibfnamefont {Gabriele}\ \bibnamefont
  {Calabrese}}, \bibinfo {author} {\bibfnamefont {Robyn}\ \bibnamefont
  {Worsley}}, \bibinfo {author} {\bibfnamefont {Subimal}\ \bibnamefont
  {Majee}}, \bibinfo {author} {\bibfnamefont {Dmitry~K.}\ \bibnamefont
  {Polyushkin}}, \bibinfo {author} {\bibfnamefont {Matthias}\ \bibnamefont
  {Paur}}, \bibinfo {author} {\bibfnamefont {Simona}\ \bibnamefont {Pace}},
  \bibinfo {author} {\bibfnamefont {Dong~Hoon}\ \bibnamefont {Keum}}, \bibinfo
  {author} {\bibfnamefont {Filippo}\ \bibnamefont {Fabbri}}, \bibinfo {author}
  {\bibfnamefont {Giuseppe}\ \bibnamefont {Iannaccone}}, \bibinfo {author}
  {\bibfnamefont {Massimo}\ \bibnamefont {Macucci}}, \bibinfo {author}
  {\bibfnamefont {Camilla}\ \bibnamefont {Coletti}}, \bibinfo {author}
  {\bibfnamefont {Thomas}\ \bibnamefont {Mueller}}, \bibinfo {author}
  {\bibfnamefont {Cinzia}\ \bibnamefont {Casiraghi}}, \ and\ \bibinfo {author}
  {\bibfnamefont {Gianluca}\ \bibnamefont {Fiori}},\ }\bibfield  {title}
  {\enquote {\bibinfo {title} {{Low-voltage 2D materials-based printed
  field-effect transistors for integrated digital and analog electronics on
  paper}},}\ }\href {\doibase 10.1038/s41467-020-17297-z} {\bibfield  {journal}
  {\bibinfo  {journal} {Nature Communications}\ }\textbf {\bibinfo {volume}
  {11}},\ \bibinfo {pages} {3566} (\bibinfo {year} {2020})}\BibitemShut
  {NoStop}%
\bibitem [{\citenamefont {Gui}\ \emph {et~al.}(2013)\citenamefont {Gui},
  \citenamefont {Zhu}, \citenamefont {Gillette}, \citenamefont {Han},
  \citenamefont {Rubloff}, \citenamefont {Hu},\ and\ \citenamefont
  {Lee}}]{gui2013natural}%
  \BibitemOpen
  \bibfield  {author} {\bibinfo {author} {\bibfnamefont {Zhe}\ \bibnamefont
  {Gui}}, \bibinfo {author} {\bibfnamefont {Hongli}\ \bibnamefont {Zhu}},
  \bibinfo {author} {\bibfnamefont {Eleanor}\ \bibnamefont {Gillette}},
  \bibinfo {author} {\bibfnamefont {Xiaogang}\ \bibnamefont {Han}}, \bibinfo
  {author} {\bibfnamefont {Gary~W.}\ \bibnamefont {Rubloff}}, \bibinfo {author}
  {\bibfnamefont {Liangbing}\ \bibnamefont {Hu}}, \ and\ \bibinfo {author}
  {\bibfnamefont {Sang~Bok}\ \bibnamefont {Lee}},\ }\bibfield  {title}
  {\enquote {\bibinfo {title} {{Natural Cellulose Fiber as Substrate for
  Supercapacitor}},}\ }\href {\doibase 10.1021/nn401818t} {\bibfield  {journal}
  {\bibinfo  {journal} {ACS Nano}\ }\textbf {\bibinfo {volume} {7}},\ \bibinfo
  {pages} {6037--6046} (\bibinfo {year} {2013})}\BibitemShut {NoStop}%
\bibitem [{\citenamefont {Brunetti}\ \emph {et~al.}(2019)\citenamefont
  {Brunetti}, \citenamefont {Operamolla}, \citenamefont {Castro-Hermosa},
  \citenamefont {Lucarelli}, \citenamefont {Manca}, \citenamefont {Farinola},\
  and\ \citenamefont {Brown}}]{brunetti2019printed}%
  \BibitemOpen
  \bibfield  {author} {\bibinfo {author} {\bibfnamefont {Francesca}\
  \bibnamefont {Brunetti}}, \bibinfo {author} {\bibfnamefont {Alessandra}\
  \bibnamefont {Operamolla}}, \bibinfo {author} {\bibfnamefont {Sergio}\
  \bibnamefont {Castro-Hermosa}}, \bibinfo {author} {\bibfnamefont {Giulia}\
  \bibnamefont {Lucarelli}}, \bibinfo {author} {\bibfnamefont {Valerio}\
  \bibnamefont {Manca}}, \bibinfo {author} {\bibfnamefont {Gianluca~M.}\
  \bibnamefont {Farinola}}, \ and\ \bibinfo {author} {\bibfnamefont
  {Thomas~M.}\ \bibnamefont {Brown}},\ }\bibfield  {title} {\enquote {\bibinfo
  {title} {{Printed Solar Cells and Energy Storage Devices on Paper
  Substrates}},}\ }\href {\doibase https://doi.org/10.1002/adfm.201806798}
  {\bibfield  {journal} {\bibinfo  {journal} {Advanced Functional Materials}\
  }\textbf {\bibinfo {volume} {29}},\ \bibinfo {pages} {1806798} (\bibinfo
  {year} {2019})}\BibitemShut {NoStop}%
\bibitem [{\citenamefont {Du}\ \emph {et~al.}(2017)\citenamefont {Du},
  \citenamefont {Zhang}, \citenamefont {Liu},\ and\ \citenamefont
  {Deng}}]{du2017nanocellulose}%
  \BibitemOpen
  \bibfield  {author} {\bibinfo {author} {\bibfnamefont {Xu}~\bibnamefont
  {Du}}, \bibinfo {author} {\bibfnamefont {Zhe}\ \bibnamefont {Zhang}},
  \bibinfo {author} {\bibfnamefont {Wei}\ \bibnamefont {Liu}}, \ and\ \bibinfo
  {author} {\bibfnamefont {Yulin}\ \bibnamefont {Deng}},\ }\bibfield  {title}
  {\enquote {\bibinfo {title} {{Nanocellulose-based conductive materials and
  their emerging applications in energy devices - A review}},}\ }\href
  {\doibase https://doi.org/10.1016/j.nanoen.2017.04.001} {\bibfield  {journal}
  {\bibinfo  {journal} {Nano Energy}\ }\textbf {\bibinfo {volume} {35}},\
  \bibinfo {pages} {299--320} (\bibinfo {year} {2017})}\BibitemShut {NoStop}%
\bibitem [{\citenamefont {Hoeng}\ \emph {et~al.}(2016)\citenamefont {Hoeng},
  \citenamefont {Denneulin},\ and\ \citenamefont {Bras}}]{hoeng2016use}%
  \BibitemOpen
  \bibfield  {author} {\bibinfo {author} {\bibfnamefont {Fanny}\ \bibnamefont
  {Hoeng}}, \bibinfo {author} {\bibfnamefont {Aurore}\ \bibnamefont
  {Denneulin}}, \ and\ \bibinfo {author} {\bibfnamefont {Julien}\ \bibnamefont
  {Bras}},\ }\bibfield  {title} {\enquote {\bibinfo {title} {{Use of
  nanocellulose in printed electronics: a review}},}\ }\href {\doibase
  10.1039/C6NR03054H} {\bibfield  {journal} {\bibinfo  {journal} {Nanoscale}\
  }\textbf {\bibinfo {volume} {8}},\ \bibinfo {pages} {13131--13154} (\bibinfo
  {year} {2016})}\BibitemShut {NoStop}%
\bibitem [{\citenamefont {Moon}\ \emph {et~al.}(2011)\citenamefont {Moon},
  \citenamefont {Martini}, \citenamefont {Nairn}, \citenamefont {Simonsen},\
  and\ \citenamefont {Youngblood}}]{moon2011cellulose}%
  \BibitemOpen
  \bibfield  {author} {\bibinfo {author} {\bibfnamefont {Robert~J.}\
  \bibnamefont {Moon}}, \bibinfo {author} {\bibfnamefont {Ashlie}\ \bibnamefont
  {Martini}}, \bibinfo {author} {\bibfnamefont {John}\ \bibnamefont {Nairn}},
  \bibinfo {author} {\bibfnamefont {John}\ \bibnamefont {Simonsen}}, \ and\
  \bibinfo {author} {\bibfnamefont {Jeff}\ \bibnamefont {Youngblood}},\
  }\bibfield  {title} {\enquote {\bibinfo {title} {{Cellulose nanomaterials
  review: structure, properties and nanocomposites}},}\ }\href {\doibase
  10.1039/C0CS00108B} {\bibfield  {journal} {\bibinfo  {journal} {Chem. Soc.
  Rev.}\ }\textbf {\bibinfo {volume} {40}},\ \bibinfo {pages} {3941--3994}
  (\bibinfo {year} {2011})}\BibitemShut {NoStop}%
\bibitem [{\citenamefont {Salas}\ \emph {et~al.}(2014)\citenamefont {Salas},
  \citenamefont {Nypelö}, \citenamefont {Rodriguez-Abreu}, \citenamefont
  {Carrillo},\ and\ \citenamefont {Rojas}}]{salas2014nanocellulose}%
  \BibitemOpen
  \bibfield  {author} {\bibinfo {author} {\bibfnamefont {Carlos}\ \bibnamefont
  {Salas}}, \bibinfo {author} {\bibfnamefont {Tiina}\ \bibnamefont {Nypelö}},
  \bibinfo {author} {\bibfnamefont {Carlos}\ \bibnamefont {Rodriguez-Abreu}},
  \bibinfo {author} {\bibfnamefont {Carlos}\ \bibnamefont {Carrillo}}, \ and\
  \bibinfo {author} {\bibfnamefont {Orlando~J.}\ \bibnamefont {Rojas}},\
  }\bibfield  {title} {\enquote {\bibinfo {title} {{Nanocellulose properties
  and applications in colloids and interfaces}},}\ }\href {\doibase
  https://doi.org/10.1016/j.cocis.2014.10.003} {\bibfield  {journal} {\bibinfo
  {journal} {Current Opinion in Colloid {\&} Interface Science}\ }\textbf
  {\bibinfo {volume} {19}},\ \bibinfo {pages} {383--396} (\bibinfo {year}
  {2014})}\BibitemShut {NoStop}%
\bibitem [{\citenamefont {Inui}\ \emph {et~al.}(2015)\citenamefont {Inui},
  \citenamefont {Koga}, \citenamefont {Nogi}, \citenamefont {Komoda},\ and\
  \citenamefont {Suganuma}}]{inui2015miniaturized}%
  \BibitemOpen
  \bibfield  {author} {\bibinfo {author} {\bibfnamefont {Tetsuji}\ \bibnamefont
  {Inui}}, \bibinfo {author} {\bibfnamefont {Hirotaka}\ \bibnamefont {Koga}},
  \bibinfo {author} {\bibfnamefont {Masaya}\ \bibnamefont {Nogi}}, \bibinfo
  {author} {\bibfnamefont {Natsuki}\ \bibnamefont {Komoda}}, \ and\ \bibinfo
  {author} {\bibfnamefont {Katsuaki}\ \bibnamefont {Suganuma}},\ }\bibfield
  {title} {\enquote {\bibinfo {title} {{A Miniaturized Flexible Antenna Printed
  on a High Dielectric Constant Nanopaper Composite}},}\ }\href {\doibase
  https://doi.org/10.1002/adma.201404555} {\bibfield  {journal} {\bibinfo
  {journal} {Advanced Materials}\ }\textbf {\bibinfo {volume} {27}},\ \bibinfo
  {pages} {1112--1116} (\bibinfo {year} {2015})}\BibitemShut {NoStop}%
\bibitem [{\citenamefont {Yang}\ \emph {et~al.}(2015)\citenamefont {Yang},
  \citenamefont {Shi}, \citenamefont {Zhitomirsky},\ and\ \citenamefont
  {Cranston}}]{yang2015cellulose}%
  \BibitemOpen
  \bibfield  {author} {\bibinfo {author} {\bibfnamefont {Xuan}\ \bibnamefont
  {Yang}}, \bibinfo {author} {\bibfnamefont {Kaiyuan}\ \bibnamefont {Shi}},
  \bibinfo {author} {\bibfnamefont {Igor}\ \bibnamefont {Zhitomirsky}}, \ and\
  \bibinfo {author} {\bibfnamefont {Emily~D.}\ \bibnamefont {Cranston}},\
  }\bibfield  {title} {\enquote {\bibinfo {title} {{Cellulose Nanocrystal
  Aerogels as Universal 3D Lightweight Substrates for Supercapacitor
  Materials}},}\ }\href {\doibase https://doi.org/10.1002/adma.201502284}
  {\bibfield  {journal} {\bibinfo  {journal} {Advanced Materials}\ }\textbf
  {\bibinfo {volume} {27}},\ \bibinfo {pages} {6104--6109} (\bibinfo {year}
  {2015})}\BibitemShut {NoStop}%
\bibitem [{\citenamefont {Fingolo}\ \emph {et~al.}(2021)\citenamefont
  {Fingolo}, \citenamefont {de~Morais}, \citenamefont {Costa}, \citenamefont
  {Corrêa}, \citenamefont {Lodi}, \citenamefont {Santhiago}, \citenamefont
  {Bernardes},\ and\ \citenamefont {Bufon}}]{fingolo2021enhanced}%
  \BibitemOpen
  \bibfield  {author} {\bibinfo {author} {\bibfnamefont {Ana~C.}\ \bibnamefont
  {Fingolo}}, \bibinfo {author} {\bibfnamefont {Vitória~B.}\ \bibnamefont
  {de~Morais}}, \bibinfo {author} {\bibfnamefont {Saionara~V.}\ \bibnamefont
  {Costa}}, \bibinfo {author} {\bibfnamefont {Cátia~C.}\ \bibnamefont
  {Corrêa}}, \bibinfo {author} {\bibfnamefont {Beatriz}\ \bibnamefont {Lodi}},
  \bibinfo {author} {\bibfnamefont {Murilo}\ \bibnamefont {Santhiago}},
  \bibinfo {author} {\bibfnamefont {Juliana~S.}\ \bibnamefont {Bernardes}}, \
  and\ \bibinfo {author} {\bibfnamefont {Carlos C.~B.}\ \bibnamefont {Bufon}},\
  }\bibfield  {title} {\enquote {\bibinfo {title} {{Enhanced Hydrophobicity in
  Nanocellulose-Based Materials: Toward Green Wearable Devices}},}\ }\href
  {\doibase 10.1021/acsabm.1c00317} {\bibfield  {journal} {\bibinfo  {journal}
  {ACS Applied Bio Materials}\ }\textbf {\bibinfo {volume} {4}},\ \bibinfo
  {pages} {6682--6689} (\bibinfo {year} {2021})}\BibitemShut {NoStop}%
\bibitem [{\citenamefont {Cao}\ \emph {et~al.}(2019)\citenamefont {Cao},
  \citenamefont {Shi}, \citenamefont {Miao}, \citenamefont {Fang},
  \citenamefont {Zhao},\ and\ \citenamefont {Feng}}]{cao2019solution}%
  \BibitemOpen
  \bibfield  {author} {\bibinfo {author} {\bibfnamefont {Shaomei}\ \bibnamefont
  {Cao}}, \bibinfo {author} {\bibfnamefont {Liyi}\ \bibnamefont {Shi}},
  \bibinfo {author} {\bibfnamefont {Miao}\ \bibnamefont {Miao}}, \bibinfo
  {author} {\bibfnamefont {Jianhui}\ \bibnamefont {Fang}}, \bibinfo {author}
  {\bibfnamefont {Hongbin}\ \bibnamefont {Zhao}}, \ and\ \bibinfo {author}
  {\bibfnamefont {Xin}\ \bibnamefont {Feng}},\ }\bibfield  {title} {\enquote
  {\bibinfo {title} {{Solution-processed flexible paper-electrode for
  lithium-ion batteries based on MoS2 nanosheets exfoliated with cellulose
  nanofibrils}},}\ }\href {\doibase
  https://doi.org/10.1016/j.electacta.2018.12.067} {\bibfield  {journal}
  {\bibinfo  {journal} {Electrochimica Acta}\ }\textbf {\bibinfo {volume}
  {298}},\ \bibinfo {pages} {22--30} (\bibinfo {year} {2019})}\BibitemShut
  {NoStop}%
\bibitem [{\citenamefont {Weng}\ \emph {et~al.}(2011)\citenamefont {Weng},
  \citenamefont {Su}, \citenamefont {Wang}, \citenamefont {Li}, \citenamefont
  {Du},\ and\ \citenamefont {Cheng}}]{weng2011graphene}%
  \BibitemOpen
  \bibfield  {author} {\bibinfo {author} {\bibfnamefont {Zhe}\ \bibnamefont
  {Weng}}, \bibinfo {author} {\bibfnamefont {Yang}\ \bibnamefont {Su}},
  \bibinfo {author} {\bibfnamefont {Da-Wei}\ \bibnamefont {Wang}}, \bibinfo
  {author} {\bibfnamefont {Feng}\ \bibnamefont {Li}}, \bibinfo {author}
  {\bibfnamefont {Jinhong}\ \bibnamefont {Du}}, \ and\ \bibinfo {author}
  {\bibfnamefont {Hui-Ming}\ \bibnamefont {Cheng}},\ }\bibfield  {title}
  {\enquote {\bibinfo {title} {{Graphene–Cellulose Paper Flexible
  Supercapacitors}},}\ }\href {\doibase https://doi.org/10.1002/aenm.201100312}
  {\bibfield  {journal} {\bibinfo  {journal} {Advanced Energy Materials}\
  }\textbf {\bibinfo {volume} {1}},\ \bibinfo {pages} {917--922} (\bibinfo
  {year} {2011})}\BibitemShut {NoStop}%
\bibitem [{\citenamefont {Xu}\ and\ \citenamefont
  {Hsieh}(2019)}]{xu2019aqueous}%
  \BibitemOpen
  \bibfield  {author} {\bibinfo {author} {\bibfnamefont {Xuezhu}\ \bibnamefont
  {Xu}}\ and\ \bibinfo {author} {\bibfnamefont {You-Lo}\ \bibnamefont
  {Hsieh}},\ }\bibfield  {title} {\enquote {\bibinfo {title} {{Aqueous
  exfoliated graphene by amphiphilic nanocellulose and its application in
  moisture-responsive foldable actuators}},}\ }\href {\doibase
  10.1039/C9NR01602C} {\bibfield  {journal} {\bibinfo  {journal} {Nanoscale}\
  }\textbf {\bibinfo {volume} {11}},\ \bibinfo {pages} {11719--11729} (\bibinfo
  {year} {2019})}\BibitemShut {NoStop}%
\bibitem [{\citenamefont {Tian}\ \emph {et~al.}(2019)\citenamefont {Tian},
  \citenamefont {VahidMohammadi}, \citenamefont {Reid}, \citenamefont {Wang},
  \citenamefont {Ouyang}, \citenamefont {Erlandsson}, \citenamefont
  {Pettersson}, \citenamefont {Wågberg}, \citenamefont {Beidaghi},\ and\
  \citenamefont {Hamedi}}]{tian2019multifunctional}%
  \BibitemOpen
  \bibfield  {author} {\bibinfo {author} {\bibfnamefont {Weiqian}\ \bibnamefont
  {Tian}}, \bibinfo {author} {\bibfnamefont {Armin}\ \bibnamefont
  {VahidMohammadi}}, \bibinfo {author} {\bibfnamefont {Michael~S.}\
  \bibnamefont {Reid}}, \bibinfo {author} {\bibfnamefont {Zhen}\ \bibnamefont
  {Wang}}, \bibinfo {author} {\bibfnamefont {Liangqi}\ \bibnamefont {Ouyang}},
  \bibinfo {author} {\bibfnamefont {Johan}\ \bibnamefont {Erlandsson}},
  \bibinfo {author} {\bibfnamefont {Torbjörn}\ \bibnamefont {Pettersson}},
  \bibinfo {author} {\bibfnamefont {Lars}\ \bibnamefont {Wågberg}}, \bibinfo
  {author} {\bibfnamefont {Majid}\ \bibnamefont {Beidaghi}}, \ and\ \bibinfo
  {author} {\bibfnamefont {Mahiar~M.}\ \bibnamefont {Hamedi}},\ }\bibfield
  {title} {\enquote {\bibinfo {title} {{Multifunctional Nanocomposites with
  High Strength and Capacitance Using 2D MXene and 1D Nanocellulose}},}\ }\href
  {\doibase https://doi.org/10.1002/adma.201902977} {\bibfield  {journal}
  {\bibinfo  {journal} {Advanced Materials}\ }\textbf {\bibinfo {volume}
  {31}},\ \bibinfo {pages} {1902977} (\bibinfo {year} {2019})}\BibitemShut
  {NoStop}%
\bibitem [{\citenamefont {Petry}\ \emph {et~al.}(2022)\citenamefont {Petry},
  \citenamefont {Silvestre}, \citenamefont {Focassio}, \citenamefont {Crasto~de
  Lima}, \citenamefont {Miwa},\ and\ \citenamefont
  {Fazzio}}]{petry2022machine}%
  \BibitemOpen
  \bibfield  {author} {\bibinfo {author} {\bibfnamefont {Romana}\ \bibnamefont
  {Petry}}, \bibinfo {author} {\bibfnamefont {Gustavo~H.}\ \bibnamefont
  {Silvestre}}, \bibinfo {author} {\bibfnamefont {Bruno}\ \bibnamefont
  {Focassio}}, \bibinfo {author} {\bibfnamefont {Felipe}\ \bibnamefont
  {Crasto~de Lima}}, \bibinfo {author} {\bibfnamefont {Roberto~H.}\
  \bibnamefont {Miwa}}, \ and\ \bibinfo {author} {\bibfnamefont {Adalberto}\
  \bibnamefont {Fazzio}},\ }\bibfield  {title} {\enquote {\bibinfo {title}
  {{Machine Learning of Microscopic Ingredients for Graphene Oxide/Cellulose
  Interaction}},}\ }\href {\doibase 10.1021/acs.langmuir.1c02780} {\bibfield
  {journal} {\bibinfo  {journal} {Langmuir}\ }\textbf {\bibinfo {volume}
  {38}},\ \bibinfo {pages} {1124--1130} (\bibinfo {year} {2022})}\BibitemShut
  {NoStop}%
\bibitem [{\citenamefont {Mianehrow}\ \emph {et~al.}(2020)\citenamefont
  {Mianehrow}, \citenamefont {Lo~Re}, \citenamefont {Carosio}, \citenamefont
  {Fina}, \citenamefont {Larsson}, \citenamefont {Chen},\ and\ \citenamefont
  {Berglund}}]{mianehrow2020strong}%
  \BibitemOpen
  \bibfield  {author} {\bibinfo {author} {\bibfnamefont {Hanieh}\ \bibnamefont
  {Mianehrow}}, \bibinfo {author} {\bibfnamefont {Giada}\ \bibnamefont
  {Lo~Re}}, \bibinfo {author} {\bibfnamefont {Federico}\ \bibnamefont
  {Carosio}}, \bibinfo {author} {\bibfnamefont {Alberto}\ \bibnamefont {Fina}},
  \bibinfo {author} {\bibfnamefont {Per~Tomas}\ \bibnamefont {Larsson}},
  \bibinfo {author} {\bibfnamefont {Pan}\ \bibnamefont {Chen}}, \ and\ \bibinfo
  {author} {\bibfnamefont {Lars~A.}\ \bibnamefont {Berglund}},\ }\bibfield
  {title} {\enquote {\bibinfo {title} {{Strong reinforcement effects in 2D
  cellulose nanofibril–graphene oxide (CNF–GO) nanocomposites due to
  GO-induced CNF ordering}},}\ }\href {\doibase 10.1039/D0TA04406G} {\bibfield
  {journal} {\bibinfo  {journal} {J. Mater. Chem. A}\ }\textbf {\bibinfo
  {volume} {8}},\ \bibinfo {pages} {17608--17620} (\bibinfo {year}
  {2020})}\BibitemShut {NoStop}%
\bibitem [{\citenamefont {Zhu}\ \emph {et~al.}(2017)\citenamefont {Zhu},
  \citenamefont {Liu},\ and\ \citenamefont {Mathew}}]{zhu2017self}%
  \BibitemOpen
  \bibfield  {author} {\bibinfo {author} {\bibfnamefont {Chuantao}\
  \bibnamefont {Zhu}}, \bibinfo {author} {\bibfnamefont {Peng}\ \bibnamefont
  {Liu}}, \ and\ \bibinfo {author} {\bibfnamefont {Aji~P.}\ \bibnamefont
  {Mathew}},\ }\bibfield  {title} {\enquote {\bibinfo {title} {{Self-Assembled
  TEMPO Cellulose Nanofibers: Graphene Oxide-Based Biohybrids for Water
  Purification}},}\ }\href {\doibase 10.1021/acsami.7b06358} {\bibfield
  {journal} {\bibinfo  {journal} {ACS Applied Materials \& Interfaces}\
  }\textbf {\bibinfo {volume} {9}},\ \bibinfo {pages} {21048--21058} (\bibinfo
  {year} {2017})}\BibitemShut {NoStop}%
\bibitem [{\citenamefont {Zhu}\ \emph {et~al.}(2022)\citenamefont {Zhu},
  \citenamefont {Wang}, \citenamefont {Sun}, \citenamefont {Fu}, \citenamefont
  {Ahmad}, \citenamefont {Fan},\ and\ \citenamefont {Gao}}]{zhu2022revealing}%
  \BibitemOpen
  \bibfield  {author} {\bibinfo {author} {\bibfnamefont {Bowen}\ \bibnamefont
  {Zhu}}, \bibinfo {author} {\bibfnamefont {Kexuan}\ \bibnamefont {Wang}},
  \bibinfo {author} {\bibfnamefont {Weisheng}\ \bibnamefont {Sun}}, \bibinfo
  {author} {\bibfnamefont {Ziyan}\ \bibnamefont {Fu}}, \bibinfo {author}
  {\bibfnamefont {Hassan}\ \bibnamefont {Ahmad}}, \bibinfo {author}
  {\bibfnamefont {Mizi}\ \bibnamefont {Fan}}, \ and\ \bibinfo {author}
  {\bibfnamefont {Haili}\ \bibnamefont {Gao}},\ }\bibfield  {title} {\enquote
  {\bibinfo {title} {{Revealing the adsorption energy and interface
  characteristic of cellulose-graphene oxide composites by first-principles
  calculations}},}\ }\href {\doibase
  https://doi.org/10.1016/j.compscitech.2021.109209} {\bibfield  {journal}
  {\bibinfo  {journal} {Composites Science and Technology}\ }\textbf {\bibinfo
  {volume} {218}},\ \bibinfo {pages} {109209} (\bibinfo {year}
  {2022})}\BibitemShut {NoStop}%
\bibitem [{\citenamefont {Perdew}\ \emph {et~al.}(1996)\citenamefont {Perdew},
  \citenamefont {Burke},\ and\ \citenamefont {Ernzerhof}}]{PBE}%
  \BibitemOpen
  \bibfield  {author} {\bibinfo {author} {\bibfnamefont {John~P.}\ \bibnamefont
  {Perdew}}, \bibinfo {author} {\bibfnamefont {Kieron}\ \bibnamefont {Burke}},
  \ and\ \bibinfo {author} {\bibfnamefont {Matthias}\ \bibnamefont
  {Ernzerhof}},\ }\bibfield  {title} {\enquote {\bibinfo {title} {{Generalized
  Gradient Approximation Made Simple}},}\ }\href {\doibase
  10.1103/PhysRevLett.77.3865} {\bibfield  {journal} {\bibinfo  {journal}
  {Phys. Rev. Lett.}\ }\textbf {\bibinfo {volume} {77}},\ \bibinfo {pages}
  {3865--3868} (\bibinfo {year} {1996})}\BibitemShut {NoStop}%
\bibitem [{\citenamefont {Bl\"ochl}(1994)}]{paw}%
  \BibitemOpen
  \bibfield  {author} {\bibinfo {author} {\bibfnamefont {P.~E.}\ \bibnamefont
  {Bl\"ochl}},\ }\bibfield  {title} {\enquote {\bibinfo {title} {{Projector
  augmented-wave method}},}\ }\href {\doibase 10.1103/PhysRevB.50.17953}
  {\bibfield  {journal} {\bibinfo  {journal} {Phys. Rev. B}\ }\textbf {\bibinfo
  {volume} {50}},\ \bibinfo {pages} {17953--17979} (\bibinfo {year}
  {1994})}\BibitemShut {NoStop}%
\bibitem [{\citenamefont {Monkhorst}\ and\ \citenamefont {Pack}(1976)}]{mp}%
  \BibitemOpen
  \bibfield  {author} {\bibinfo {author} {\bibfnamefont {Hendrik~J.}\
  \bibnamefont {Monkhorst}}\ and\ \bibinfo {author} {\bibfnamefont {James~D.}\
  \bibnamefont {Pack}},\ }\bibfield  {title} {\enquote {\bibinfo {title}
  {{Special points for Brillouin-zone integrations}},}\ }\href {\doibase
  10.1103/PhysRevB.13.5188} {\bibfield  {journal} {\bibinfo  {journal} {Phys.
  Rev. B}\ }\textbf {\bibinfo {volume} {13}},\ \bibinfo {pages} {5188--5192}
  (\bibinfo {year} {1976})}\BibitemShut {NoStop}%
\bibitem [{\citenamefont {Kresse}\ and\ \citenamefont
  {Furthmüller}(1996)}]{vasp1}%
  \BibitemOpen
  \bibfield  {author} {\bibinfo {author} {\bibfnamefont {G.}~\bibnamefont
  {Kresse}}\ and\ \bibinfo {author} {\bibfnamefont {J.}~\bibnamefont
  {Furthmüller}},\ }\bibfield  {title} {\enquote {\bibinfo {title}
  {{Efficiency of ab-initio total energy calculations for metals and
  semiconductors using a plane-wave basis set}},}\ }\href {\doibase
  https://doi.org/10.1016/0927-0256(96)00008-0} {\bibfield  {journal} {\bibinfo
   {journal} {Computational Materials Science}\ }\textbf {\bibinfo {volume}
  {6}},\ \bibinfo {pages} {15--50} (\bibinfo {year} {1996})}\BibitemShut
  {NoStop}%
\bibitem [{\citenamefont {Kresse}\ and\ \citenamefont
  {Furthm\"uller}(1996)}]{vasp2}%
  \BibitemOpen
  \bibfield  {author} {\bibinfo {author} {\bibfnamefont {G.}~\bibnamefont
  {Kresse}}\ and\ \bibinfo {author} {\bibfnamefont {J.}~\bibnamefont
  {Furthm\"uller}},\ }\bibfield  {title} {\enquote {\bibinfo {title}
  {{Efficient iterative schemes for ab initio total-energy calculations using a
  plane-wave basis set}},}\ }\href {\doibase 10.1103/PhysRevB.54.11169}
  {\bibfield  {journal} {\bibinfo  {journal} {Phys. Rev. B}\ }\textbf {\bibinfo
  {volume} {54}},\ \bibinfo {pages} {11169--11186} (\bibinfo {year}
  {1996})}\BibitemShut {NoStop}%
\bibitem [{\citenamefont {Mathew}\ \emph {et~al.}(2018)\citenamefont {Mathew},
  \citenamefont {Kolluru},\ and\ \citenamefont {Hennig}}]{VASPsol-Software}%
  \BibitemOpen
  \bibfield  {author} {\bibinfo {author} {\bibfnamefont {K.}~\bibnamefont
  {Mathew}}, \bibinfo {author} {\bibfnamefont {V.~S.~Chaitanya}\ \bibnamefont
  {Kolluru}}, \ and\ \bibinfo {author} {\bibfnamefont {R.~G.}\ \bibnamefont
  {Hennig}},\ }\bibfield  {title} {\enquote {\bibinfo {title} {{VASPsol:
  Implicit solvation and electrolyte model for density-functional theory}},}\
  }\href {\doibase 10.5281/zenodo.2555053} {\bibfield  {journal} {\bibinfo
  {journal} {GitHub repository}\ } (\bibinfo {year} {2018}),\
  10.5281/zenodo.2555053}\BibitemShut {NoStop}%
\bibitem [{\citenamefont {Mathew}\ \emph {et~al.}(2014)\citenamefont {Mathew},
  \citenamefont {Sundararaman}, \citenamefont {Letchworth-Weaver},
  \citenamefont {Arias},\ and\ \citenamefont {Hennig}}]{mathew2014implicit}%
  \BibitemOpen
  \bibfield  {author} {\bibinfo {author} {\bibfnamefont {Kiran}\ \bibnamefont
  {Mathew}}, \bibinfo {author} {\bibfnamefont {Ravishankar}\ \bibnamefont
  {Sundararaman}}, \bibinfo {author} {\bibfnamefont {Kendra}\ \bibnamefont
  {Letchworth-Weaver}}, \bibinfo {author} {\bibfnamefont {T.~A.}\ \bibnamefont
  {Arias}}, \ and\ \bibinfo {author} {\bibfnamefont {Richard~G.}\ \bibnamefont
  {Hennig}},\ }\bibfield  {title} {\enquote {\bibinfo {title} {{Implicit
  solvation model for density-functional study of nanocrystal surfaces and
  reaction pathways}},}\ }\href {\doibase 10.1063/1.4865107} {\bibfield
  {journal} {\bibinfo  {journal} {The Journal of Chemical Physics}\ }\textbf
  {\bibinfo {volume} {140}},\ \bibinfo {pages} {084106} (\bibinfo {year}
  {2014})}\BibitemShut {NoStop}%
\bibitem [{\citenamefont {Mathew}\ \emph {et~al.}(2019)\citenamefont {Mathew},
  \citenamefont {Kolluru}, \citenamefont {Mula}, \citenamefont {Steinmann},\
  and\ \citenamefont {Hennig}}]{mathew2019implicit}%
  \BibitemOpen
  \bibfield  {author} {\bibinfo {author} {\bibfnamefont {Kiran}\ \bibnamefont
  {Mathew}}, \bibinfo {author} {\bibfnamefont {V.~S.~Chaitanya}\ \bibnamefont
  {Kolluru}}, \bibinfo {author} {\bibfnamefont {Srinidhi}\ \bibnamefont
  {Mula}}, \bibinfo {author} {\bibfnamefont {Stephan~N.}\ \bibnamefont
  {Steinmann}}, \ and\ \bibinfo {author} {\bibfnamefont {Richard~G.}\
  \bibnamefont {Hennig}},\ }\bibfield  {title} {\enquote {\bibinfo {title}
  {{Implicit self-consistent electrolyte model in plane-wave density-functional
  theory}},}\ }\href {\doibase 10.1063/1.5132354} {\bibfield  {journal}
  {\bibinfo  {journal} {The Journal of Chemical Physics}\ }\textbf {\bibinfo
  {volume} {151}},\ \bibinfo {pages} {234101} (\bibinfo {year}
  {2019})}\BibitemShut {NoStop}%
\bibitem [{\citenamefont {Thonhauser}\ \emph {et~al.}(2015)\citenamefont
  {Thonhauser}, \citenamefont {Zuluaga}, \citenamefont {Arter}, \citenamefont
  {Berland}, \citenamefont {Schr\"oder},\ and\ \citenamefont
  {Hyldgaard}}]{thonhauserPRL2015}%
  \BibitemOpen
  \bibfield  {author} {\bibinfo {author} {\bibfnamefont {T.}~\bibnamefont
  {Thonhauser}}, \bibinfo {author} {\bibfnamefont {S.}~\bibnamefont {Zuluaga}},
  \bibinfo {author} {\bibfnamefont {C.~A.}\ \bibnamefont {Arter}}, \bibinfo
  {author} {\bibfnamefont {K.}~\bibnamefont {Berland}}, \bibinfo {author}
  {\bibfnamefont {E.}~\bibnamefont {Schr\"oder}}, \ and\ \bibinfo {author}
  {\bibfnamefont {P.}~\bibnamefont {Hyldgaard}},\ }\bibfield  {title} {\enquote
  {\bibinfo {title} {{Spin Signature of Nonlocal Correlation Binding in
  Metal-Organic Frameworks}},}\ }\href {\doibase
  10.1103/PhysRevLett.115.136402} {\bibfield  {journal} {\bibinfo  {journal}
  {Phys. Rev. Lett.}\ }\textbf {\bibinfo {volume} {115}},\ \bibinfo {pages}
  {136402} (\bibinfo {year} {2015})}\BibitemShut {NoStop}%
\bibitem [{\citenamefont {Thonhauser}\ \emph {et~al.}(2007)\citenamefont
  {Thonhauser}, \citenamefont {Cooper}, \citenamefont {Li}, \citenamefont
  {Puzder}, \citenamefont {Hyldgaard},\ and\ \citenamefont
  {Langreth}}]{thonhauserPRB2007}%
  \BibitemOpen
  \bibfield  {author} {\bibinfo {author} {\bibfnamefont {T.}~\bibnamefont
  {Thonhauser}}, \bibinfo {author} {\bibfnamefont {Valentino~R.}\ \bibnamefont
  {Cooper}}, \bibinfo {author} {\bibfnamefont {Shen}\ \bibnamefont {Li}},
  \bibinfo {author} {\bibfnamefont {Aaron}\ \bibnamefont {Puzder}}, \bibinfo
  {author} {\bibfnamefont {Per}\ \bibnamefont {Hyldgaard}}, \ and\ \bibinfo
  {author} {\bibfnamefont {David~C.}\ \bibnamefont {Langreth}},\ }\bibfield
  {title} {\enquote {\bibinfo {title} {{Van der Waals density functional:
  Self-consistent potential and the nature of the van der Waals bond}},}\
  }\href {\doibase 10.1103/PhysRevB.76.125112} {\bibfield  {journal} {\bibinfo
  {journal} {Phys. Rev. B}\ }\textbf {\bibinfo {volume} {76}},\ \bibinfo
  {pages} {125112} (\bibinfo {year} {2007})}\BibitemShut {NoStop}%
\bibitem [{\citenamefont {Langreth}\ \emph {et~al.}(2009)\citenamefont
  {Langreth}, \citenamefont {Lundqvist}, \citenamefont {Chakarova-Käck},
  \citenamefont {Cooper}, \citenamefont {Dion}, \citenamefont {Hyldgaard},
  \citenamefont {Kelkkanen}, \citenamefont {Kleis}, \citenamefont {Kong},
  \citenamefont {Li}, \citenamefont {Moses}, \citenamefont {Murray},
  \citenamefont {Puzder}, \citenamefont {Rydberg}, \citenamefont {Schröder},\
  and\ \citenamefont {Thonhauser}}]{langrethJPhysC2009}%
  \BibitemOpen
  \bibfield  {author} {\bibinfo {author} {\bibfnamefont {D~C}\ \bibnamefont
  {Langreth}}, \bibinfo {author} {\bibfnamefont {B~I}\ \bibnamefont
  {Lundqvist}}, \bibinfo {author} {\bibfnamefont {S~D}\ \bibnamefont
  {Chakarova-Käck}}, \bibinfo {author} {\bibfnamefont {V~R}\ \bibnamefont
  {Cooper}}, \bibinfo {author} {\bibfnamefont {M}~\bibnamefont {Dion}},
  \bibinfo {author} {\bibfnamefont {P}~\bibnamefont {Hyldgaard}}, \bibinfo
  {author} {\bibfnamefont {A}~\bibnamefont {Kelkkanen}}, \bibinfo {author}
  {\bibfnamefont {J}~\bibnamefont {Kleis}}, \bibinfo {author} {\bibfnamefont
  {Lingzhu}\ \bibnamefont {Kong}}, \bibinfo {author} {\bibfnamefont {Shen}\
  \bibnamefont {Li}}, \bibinfo {author} {\bibfnamefont {P~G}\ \bibnamefont
  {Moses}}, \bibinfo {author} {\bibfnamefont {E}~\bibnamefont {Murray}},
  \bibinfo {author} {\bibfnamefont {A}~\bibnamefont {Puzder}}, \bibinfo
  {author} {\bibfnamefont {H}~\bibnamefont {Rydberg}}, \bibinfo {author}
  {\bibfnamefont {E}~\bibnamefont {Schröder}}, \ and\ \bibinfo {author}
  {\bibfnamefont {T}~\bibnamefont {Thonhauser}},\ }\bibfield  {title} {\enquote
  {\bibinfo {title} {{A density functional for sparse matter}},}\ }\href
  {\doibase 10.1088/0953-8984/21/8/084203} {\bibfield  {journal} {\bibinfo
  {journal} {Journal of Physics: Condensed Matter}\ }\textbf {\bibinfo {volume}
  {21}},\ \bibinfo {pages} {084203} (\bibinfo {year} {2009})}\BibitemShut
  {NoStop}%
\bibitem [{\citenamefont {Klime\v{s}}\ \emph {et~al.}(2009)\citenamefont
  {Klime\v{s}}, \citenamefont {Bowler},\ and\ \citenamefont
  {Michaelides}}]{klimesJPhysC2010}%
  \BibitemOpen
  \bibfield  {author} {\bibinfo {author} {\bibfnamefont {Ji\v{r}\'{\i}}\
  \bibnamefont {Klime\v{s}}}, \bibinfo {author} {\bibfnamefont {David~R}\
  \bibnamefont {Bowler}}, \ and\ \bibinfo {author} {\bibfnamefont {Angelos}\
  \bibnamefont {Michaelides}},\ }\bibfield  {title} {\enquote {\bibinfo {title}
  {{Chemical accuracy for the van der Waals density functional}},}\ }\href
  {\doibase 10.1088/0953-8984/22/2/022201} {\bibfield  {journal} {\bibinfo
  {journal} {Journal of Physics: Condensed Matter}\ }\textbf {\bibinfo {volume}
  {22}},\ \bibinfo {pages} {022201} (\bibinfo {year} {2009})}\BibitemShut
  {NoStop}%
\bibitem [{\citenamefont {Klime\ifmmode~\check{s}\else \v{s}\fi{}}\ \emph
  {et~al.}(2011)\citenamefont {Klime\ifmmode~\check{s}\else \v{s}\fi{}},
  \citenamefont {Bowler},\ and\ \citenamefont {Michaelides}}]{klimevsPRB2011}%
  \BibitemOpen
  \bibfield  {author} {\bibinfo {author} {\bibfnamefont {Ji\ifmmode
  \check{r}\else~\v{r}\fi{}\'{\i}}\ \bibnamefont {Klime\ifmmode~\check{s}\else
  \v{s}\fi{}}}, \bibinfo {author} {\bibfnamefont {David~R.}\ \bibnamefont
  {Bowler}}, \ and\ \bibinfo {author} {\bibfnamefont {Angelos}\ \bibnamefont
  {Michaelides}},\ }\bibfield  {title} {\enquote {\bibinfo {title} {{Van der
  Waals density functionals applied to solids}},}\ }\href {\doibase
  10.1103/PhysRevB.83.195131} {\bibfield  {journal} {\bibinfo  {journal} {Phys.
  Rev. B}\ }\textbf {\bibinfo {volume} {83}},\ \bibinfo {pages} {195131}
  (\bibinfo {year} {2011})}\BibitemShut {NoStop}%
\bibitem [{\citenamefont {Bun\ifmmode~\u{a}\else \u{a}\fi{}u}\ and\
  \citenamefont {Calandra}(2013)}]{xas1}%
  \BibitemOpen
  \bibfield  {author} {\bibinfo {author} {\bibfnamefont {Oana}\ \bibnamefont
  {Bun\ifmmode~\u{a}\else \u{a}\fi{}u}}\ and\ \bibinfo {author} {\bibfnamefont
  {Matteo}\ \bibnamefont {Calandra}},\ }\bibfield  {title} {\enquote {\bibinfo
  {title} {"projector augmented wave calculation of x-ray absorption spectra at
  the l2,3 edges"},}\ }\href {\doibase 10.1103/PhysRevB.87.205105} {\bibfield
  {journal} {\bibinfo  {journal} {Phys. Rev. B}\ }\textbf {\bibinfo {volume}
  {87}},\ \bibinfo {pages} {205105} (\bibinfo {year} {2013})}\BibitemShut
  {NoStop}%
\bibitem [{\citenamefont {Gougoussis}\ \emph {et~al.}(2009)\citenamefont
  {Gougoussis}, \citenamefont {Calandra}, \citenamefont {Seitsonen},\ and\
  \citenamefont {Mauri}}]{xas2}%
  \BibitemOpen
  \bibfield  {author} {\bibinfo {author} {\bibfnamefont {Christos}\
  \bibnamefont {Gougoussis}}, \bibinfo {author} {\bibfnamefont {Matteo}\
  \bibnamefont {Calandra}}, \bibinfo {author} {\bibfnamefont {Ari~P.}\
  \bibnamefont {Seitsonen}}, \ and\ \bibinfo {author} {\bibfnamefont
  {Francesco}\ \bibnamefont {Mauri}},\ }\bibfield  {title} {\enquote {\bibinfo
  {title} {{First-principles calculations of x-ray absorption in a scheme based
  on ultrasoft pseudopotentials: From $\alpha$-quartz to high-$T_c$
  compounds}},}\ }\href {\doibase 10.1103/PhysRevB.80.075102} {\bibfield
  {journal} {\bibinfo  {journal} {Phys. Rev. B}\ }\textbf {\bibinfo {volume}
  {80}},\ \bibinfo {pages} {075102} (\bibinfo {year} {2009})}\BibitemShut
  {NoStop}%
\bibitem [{\citenamefont {Taillefumier}\ \emph {et~al.}(2002)\citenamefont
  {Taillefumier}, \citenamefont {Cabaret}, \citenamefont {Flank},\ and\
  \citenamefont {Mauri}}]{xas3}%
  \BibitemOpen
  \bibfield  {author} {\bibinfo {author} {\bibfnamefont {Mathieu}\ \bibnamefont
  {Taillefumier}}, \bibinfo {author} {\bibfnamefont {Delphine}\ \bibnamefont
  {Cabaret}}, \bibinfo {author} {\bibfnamefont {Anne-Marie}\ \bibnamefont
  {Flank}}, \ and\ \bibinfo {author} {\bibfnamefont {Francesco}\ \bibnamefont
  {Mauri}},\ }\bibfield  {title} {\enquote {\bibinfo {title} {{X-ray absorption
  near-edge structure calculations with the pseudopotentials: Application to
  the K edge in diamond and $\alpha$-quartz}},}\ }\href {\doibase
  10.1103/PhysRevB.66.195107} {\bibfield  {journal} {\bibinfo  {journal} {Phys.
  Rev. B}\ }\textbf {\bibinfo {volume} {66}},\ \bibinfo {pages} {195107}
  (\bibinfo {year} {2002})}\BibitemShut {NoStop}%
\bibitem [{\citenamefont {Giannozzi}\ \emph {et~al.}(2009)\citenamefont
  {Giannozzi}, \citenamefont {Baroni}, \citenamefont {Bonini}, \citenamefont
  {Calandra}, \citenamefont {Car}, \citenamefont {Cavazzoni}, \citenamefont
  {Ceresoli}, \citenamefont {Chiarotti}, \citenamefont {Cococcioni},
  \citenamefont {Dabo}, \citenamefont {Corso}, \citenamefont {de~Gironcoli},
  \citenamefont {Fabris}, \citenamefont {Fratesi}, \citenamefont {Gebauer},
  \citenamefont {Gerstmann}, \citenamefont {Gougoussis}, \citenamefont
  {Kokalj}, \citenamefont {Lazzeri}, \citenamefont {Martin-Samos},
  \citenamefont {Marzari}, \citenamefont {Mauri}, \citenamefont {Mazzarello},
  \citenamefont {Paolini}, \citenamefont {Pasquarello}, \citenamefont
  {Paulatto}, \citenamefont {Sbraccia}, \citenamefont {Scandolo}, \citenamefont
  {Sclauzero}, \citenamefont {Seitsonen}, \citenamefont {Smogunov},
  \citenamefont {Umari},\ and\ \citenamefont {Wentzcovitch}}]{espresso}%
  \BibitemOpen
  \bibfield  {author} {\bibinfo {author} {\bibfnamefont {Paolo}\ \bibnamefont
  {Giannozzi}}, \bibinfo {author} {\bibfnamefont {Stefano}\ \bibnamefont
  {Baroni}}, \bibinfo {author} {\bibfnamefont {Nicola}\ \bibnamefont {Bonini}},
  \bibinfo {author} {\bibfnamefont {Matteo}\ \bibnamefont {Calandra}}, \bibinfo
  {author} {\bibfnamefont {Roberto}\ \bibnamefont {Car}}, \bibinfo {author}
  {\bibfnamefont {Carlo}\ \bibnamefont {Cavazzoni}}, \bibinfo {author}
  {\bibfnamefont {Davide}\ \bibnamefont {Ceresoli}}, \bibinfo {author}
  {\bibfnamefont {Guido~L}\ \bibnamefont {Chiarotti}}, \bibinfo {author}
  {\bibfnamefont {Matteo}\ \bibnamefont {Cococcioni}}, \bibinfo {author}
  {\bibfnamefont {Ismaila}\ \bibnamefont {Dabo}}, \bibinfo {author}
  {\bibfnamefont {Andrea~Dal}\ \bibnamefont {Corso}}, \bibinfo {author}
  {\bibfnamefont {Stefano}\ \bibnamefont {de~Gironcoli}}, \bibinfo {author}
  {\bibfnamefont {Stefano}\ \bibnamefont {Fabris}}, \bibinfo {author}
  {\bibfnamefont {Guido}\ \bibnamefont {Fratesi}}, \bibinfo {author}
  {\bibfnamefont {Ralph}\ \bibnamefont {Gebauer}}, \bibinfo {author}
  {\bibfnamefont {Uwe}\ \bibnamefont {Gerstmann}}, \bibinfo {author}
  {\bibfnamefont {Christos}\ \bibnamefont {Gougoussis}}, \bibinfo {author}
  {\bibfnamefont {Anton}\ \bibnamefont {Kokalj}}, \bibinfo {author}
  {\bibfnamefont {Michele}\ \bibnamefont {Lazzeri}}, \bibinfo {author}
  {\bibfnamefont {Layla}\ \bibnamefont {Martin-Samos}}, \bibinfo {author}
  {\bibfnamefont {Nicola}\ \bibnamefont {Marzari}}, \bibinfo {author}
  {\bibfnamefont {Francesco}\ \bibnamefont {Mauri}}, \bibinfo {author}
  {\bibfnamefont {Riccardo}\ \bibnamefont {Mazzarello}}, \bibinfo {author}
  {\bibfnamefont {Stefano}\ \bibnamefont {Paolini}}, \bibinfo {author}
  {\bibfnamefont {Alfredo}\ \bibnamefont {Pasquarello}}, \bibinfo {author}
  {\bibfnamefont {Lorenzo}\ \bibnamefont {Paulatto}}, \bibinfo {author}
  {\bibfnamefont {Carlo}\ \bibnamefont {Sbraccia}}, \bibinfo {author}
  {\bibfnamefont {Sandro}\ \bibnamefont {Scandolo}}, \bibinfo {author}
  {\bibfnamefont {Gabriele}\ \bibnamefont {Sclauzero}}, \bibinfo {author}
  {\bibfnamefont {Ari~P}\ \bibnamefont {Seitsonen}}, \bibinfo {author}
  {\bibfnamefont {Alexander}\ \bibnamefont {Smogunov}}, \bibinfo {author}
  {\bibfnamefont {Paolo}\ \bibnamefont {Umari}}, \ and\ \bibinfo {author}
  {\bibfnamefont {Renata~M}\ \bibnamefont {Wentzcovitch}},\ }\bibfield  {title}
  {\enquote {\bibinfo {title} {{QUANTUM ESPRESSO: a modular and open-source
  software project for quantum simulations of materials}},}\ }\href {\doibase
  10.1088/0953-8984/21/39/395502} {\bibfield  {journal} {\bibinfo  {journal}
  {Journal of Physics: Condensed Matter}\ }\textbf {\bibinfo {volume} {21}},\
  \bibinfo {pages} {395502} (\bibinfo {year} {2009})}\BibitemShut {NoStop}%
\bibitem [{\citenamefont {Berland}\ \emph {et~al.}(2015)\citenamefont
  {Berland}, \citenamefont {Cooper}, \citenamefont {Lee}, \citenamefont
  {Schröder}, \citenamefont {Thonhauser}, \citenamefont {Hyldgaard},\ and\
  \citenamefont {Lundqvist}}]{berland2015van}%
  \BibitemOpen
  \bibfield  {author} {\bibinfo {author} {\bibfnamefont {Kristian}\
  \bibnamefont {Berland}}, \bibinfo {author} {\bibfnamefont {Valentino~R}\
  \bibnamefont {Cooper}}, \bibinfo {author} {\bibfnamefont {Kyuho}\
  \bibnamefont {Lee}}, \bibinfo {author} {\bibfnamefont {Elsebeth}\
  \bibnamefont {Schröder}}, \bibinfo {author} {\bibfnamefont {T}~\bibnamefont
  {Thonhauser}}, \bibinfo {author} {\bibfnamefont {Per}\ \bibnamefont
  {Hyldgaard}}, \ and\ \bibinfo {author} {\bibfnamefont {Bengt~I}\ \bibnamefont
  {Lundqvist}},\ }\bibfield  {title} {\enquote {\bibinfo {title} {{van der
  Waals forces in density functional theory: a review of the vdW-DF method}},}\
  }\href {\doibase 10.1088/0034-4885/78/6/066501} {\bibfield  {journal}
  {\bibinfo  {journal} {Reports on Progress in Physics}\ }\textbf {\bibinfo
  {volume} {78}},\ \bibinfo {pages} {066501} (\bibinfo {year}
  {2015})}\BibitemShut {NoStop}%
\bibitem [{\citenamefont {Pickard}\ and\ \citenamefont {Mauri}(2001)}]{gipaw}%
  \BibitemOpen
  \bibfield  {author} {\bibinfo {author} {\bibfnamefont {Chris~J.}\
  \bibnamefont {Pickard}}\ and\ \bibinfo {author} {\bibfnamefont {Francesco}\
  \bibnamefont {Mauri}},\ }\bibfield  {title} {\enquote {\bibinfo {title}
  {{All-electron magnetic response with pseudopotentials: NMR chemical
  shifts}},}\ }\href {\doibase 10.1103/PhysRevB.63.245101} {\bibfield
  {journal} {\bibinfo  {journal} {Phys. Rev. B}\ }\textbf {\bibinfo {volume}
  {63}},\ \bibinfo {pages} {245101} (\bibinfo {year} {2001})}\BibitemShut
  {NoStop}%
\bibitem [{\citenamefont {Fan}\ \emph {et~al.}(2011)\citenamefont {Fan},
  \citenamefont {Zhao}, \citenamefont {Wang}, \citenamefont {Zhang},\ and\
  \citenamefont {Zhang}}]{fan2011tunable}%
  \BibitemOpen
  \bibfield  {author} {\bibinfo {author} {\bibfnamefont {Yingcai}\ \bibnamefont
  {Fan}}, \bibinfo {author} {\bibfnamefont {Mingwen}\ \bibnamefont {Zhao}},
  \bibinfo {author} {\bibfnamefont {Zhenhai}\ \bibnamefont {Wang}}, \bibinfo
  {author} {\bibfnamefont {Xuejuan}\ \bibnamefont {Zhang}}, \ and\ \bibinfo
  {author} {\bibfnamefont {Hongyu}\ \bibnamefont {Zhang}},\ }\bibfield  {title}
  {\enquote {\bibinfo {title} {{Tunable electronic structures of graphene/boron
  nitride heterobilayers}},}\ }\href {\doibase 10.1063/1.3556640} {\bibfield
  {journal} {\bibinfo  {journal} {Applied Physics Letters}\ }\textbf {\bibinfo
  {volume} {98}},\ \bibinfo {pages} {083103} (\bibinfo {year}
  {2011})}\BibitemShut {NoStop}%
\bibitem [{\citenamefont {Silvestre}\ \emph {et~al.}(2021)\citenamefont
  {Silvestre}, \citenamefont {Pinto}, \citenamefont {Bernardes}, \citenamefont
  {Miwa},\ and\ \citenamefont {Fazzio}}]{silvestreJPhysChemB2021}%
  \BibitemOpen
  \bibfield  {author} {\bibinfo {author} {\bibfnamefont {Gustavo~H.}\
  \bibnamefont {Silvestre}}, \bibinfo {author} {\bibfnamefont {Lidiane~O.}\
  \bibnamefont {Pinto}}, \bibinfo {author} {\bibfnamefont {Juliana~S.}\
  \bibnamefont {Bernardes}}, \bibinfo {author} {\bibfnamefont {Roberto~H.}\
  \bibnamefont {Miwa}}, \ and\ \bibinfo {author} {\bibfnamefont {Adalberto}\
  \bibnamefont {Fazzio}},\ }\bibfield  {title} {\enquote {\bibinfo {title}
  {{Disassembly of TEMPO-Oxidized Cellulose Fibers: Intersheet and Interchain
  Interactions in the Isolation of Nanofibers and Unitary Chains}},}\ }\href
  {\doibase 10.1021/acs.jpcb.1c01928} {\bibfield  {journal} {\bibinfo
  {journal} {The Journal of Physical Chemistry B}\ }\textbf {\bibinfo {volume}
  {125}},\ \bibinfo {pages} {3717--3724} (\bibinfo {year} {2021})}\BibitemShut
  {NoStop}%
\bibitem [{\citenamefont {Alqus}\ \emph {et~al.}(2015)\citenamefont {Alqus},
  \citenamefont {Eichhorn},\ and\ \citenamefont {Bryce}}]{alqusBiomacro2015}%
  \BibitemOpen
  \bibfield  {author} {\bibinfo {author} {\bibfnamefont {Rasha}\ \bibnamefont
  {Alqus}}, \bibinfo {author} {\bibfnamefont {Stephen~J.}\ \bibnamefont
  {Eichhorn}}, \ and\ \bibinfo {author} {\bibfnamefont {Richard~A.}\
  \bibnamefont {Bryce}},\ }\bibfield  {title} {\enquote {\bibinfo {title}
  {{Molecular Dynamics of Cellulose Amphiphilicity at the Graphene–Water
  Interface}},}\ }\href {\doibase 10.1021/acs.biomac.5b00307} {\bibfield
  {journal} {\bibinfo  {journal} {Biomacromolecules}\ }\textbf {\bibinfo
  {volume} {16}},\ \bibinfo {pages} {1771--1783} (\bibinfo {year}
  {2015})}\BibitemShut {NoStop}%
\bibitem [{\citenamefont {Mianehrow}\ \emph {et~al.}(2022)\citenamefont
  {Mianehrow}, \citenamefont {Berglund},\ and\ \citenamefont
  {Wohlert}}]{mianehrow2022interface}%
  \BibitemOpen
  \bibfield  {author} {\bibinfo {author} {\bibfnamefont {Hanieh}\ \bibnamefont
  {Mianehrow}}, \bibinfo {author} {\bibfnamefont {Lars~A.}\ \bibnamefont
  {Berglund}}, \ and\ \bibinfo {author} {\bibfnamefont {Jakob}\ \bibnamefont
  {Wohlert}},\ }\bibfield  {title} {\enquote {\bibinfo {title} {{Interface
  effects from moisture in nanocomposites of 2D graphene oxide in cellulose
  nanofiber (CNF) matrix – A molecular dynamics study}},}\ }\href {\doibase
  10.1039/D1TA09286C} {\bibfield  {journal} {\bibinfo  {journal} {J. Mater.
  Chem. A}\ }\textbf {\bibinfo {volume} {10}},\ \bibinfo {pages} {2122--2132}
  (\bibinfo {year} {2022})}\BibitemShut {NoStop}%
\bibitem [{Note1()}]{Note1}%
  \BibitemOpen
  \bibinfo {note} {$E^s$ is defined as the diference between the vacuum total
  energy and solvent total energy.}\BibitemShut {Stop}%
\bibitem [{\citenamefont {Li}\ \emph {et~al.}(2011)\citenamefont {Li},
  \citenamefont {Lin},\ and\ \citenamefont {Davenport}}]{liJPhysChemC2011}%
  \BibitemOpen
  \bibfield  {author} {\bibinfo {author} {\bibfnamefont {Yan}\ \bibnamefont
  {Li}}, \bibinfo {author} {\bibfnamefont {Milo}\ \bibnamefont {Lin}}, \ and\
  \bibinfo {author} {\bibfnamefont {James~W.}\ \bibnamefont {Davenport}},\
  }\bibfield  {title} {\enquote {\bibinfo {title} {{Ab Initio Studies of
  Cellulose I: Crystal Structure, Intermolecular Forces, and Interactions with
  Water}},}\ }\href {\doibase 10.1021/jp2006759} {\bibfield  {journal}
  {\bibinfo  {journal} {The Journal of Physical Chemistry C}\ }\textbf
  {\bibinfo {volume} {115}},\ \bibinfo {pages} {11533--11539} (\bibinfo {year}
  {2011})}\BibitemShut {NoStop}%
\bibitem [{\citenamefont {Prendergast}\ and\ \citenamefont
  {Galli}(2006)}]{prendergast2006x}%
  \BibitemOpen
  \bibfield  {author} {\bibinfo {author} {\bibfnamefont {David}\ \bibnamefont
  {Prendergast}}\ and\ \bibinfo {author} {\bibfnamefont {Giulia}\ \bibnamefont
  {Galli}},\ }\bibfield  {title} {\enquote {\bibinfo {title} {{X-Ray Absorption
  Spectra of Water from First Principles Calculations}},}\ }\href {\doibase
  10.1103/PhysRevLett.96.215502} {\bibfield  {journal} {\bibinfo  {journal}
  {Phys. Rev. Lett.}\ }\textbf {\bibinfo {volume} {96}},\ \bibinfo {pages}
  {215502} (\bibinfo {year} {2006})}\BibitemShut {NoStop}%
\bibitem [{\citenamefont {Crasto~de Lima}\ \emph {et~al.}(2020)\citenamefont
  {Crasto~de Lima}, \citenamefont {Fazzio}, \citenamefont {McLean},\ and\
  \citenamefont {Miwa}}]{de2020simulations}%
  \BibitemOpen
  \bibfield  {author} {\bibinfo {author} {\bibfnamefont {F.}~\bibnamefont
  {Crasto~de Lima}}, \bibinfo {author} {\bibfnamefont {A.}~\bibnamefont
  {Fazzio}}, \bibinfo {author} {\bibfnamefont {A.~B.}\ \bibnamefont {McLean}},
  \ and\ \bibinfo {author} {\bibfnamefont {R.~H.}\ \bibnamefont {Miwa}},\
  }\bibfield  {title} {\enquote {\bibinfo {title} {{Simulations of X-ray
  absorption spectroscopy and energetic conformation of N-heterocyclic carbenes
  on Au(111)}},}\ }\href {\doibase 10.1039/D0CP04240D} {\bibfield  {journal}
  {\bibinfo  {journal} {Phys. Chem. Chem. Phys.}\ }\textbf {\bibinfo {volume}
  {22}},\ \bibinfo {pages} {21504--21511} (\bibinfo {year} {2020})}\BibitemShut
  {NoStop}%
\bibitem [{\citenamefont {Inayeh}\ \emph {et~al.}(2021)\citenamefont {Inayeh},
  \citenamefont {Groome}, \citenamefont {Singh}, \citenamefont {Veinot},
  \citenamefont {de~Lima}, \citenamefont {Miwa}, \citenamefont {Crudden},\ and\
  \citenamefont {McLean}}]{inayeh2021self}%
  \BibitemOpen
  \bibfield  {author} {\bibinfo {author} {\bibfnamefont {Alex}\ \bibnamefont
  {Inayeh}}, \bibinfo {author} {\bibfnamefont {Ryan R.~K.}\ \bibnamefont
  {Groome}}, \bibinfo {author} {\bibfnamefont {Ishwar}\ \bibnamefont {Singh}},
  \bibinfo {author} {\bibfnamefont {Alex~J.}\ \bibnamefont {Veinot}}, \bibinfo
  {author} {\bibfnamefont {Felipe~Crasto}\ \bibnamefont {de~Lima}}, \bibinfo
  {author} {\bibfnamefont {Roberto~H.}\ \bibnamefont {Miwa}}, \bibinfo {author}
  {\bibfnamefont {Cathleen~M.}\ \bibnamefont {Crudden}}, \ and\ \bibinfo
  {author} {\bibfnamefont {Alastair~B.}\ \bibnamefont {McLean}},\ }\bibfield
  {title} {\enquote {\bibinfo {title} {{Self-assembly of N-heterocyclic
  carbenes on Au(111)}},}\ }\href {\doibase 10.1038/s41467-021-23940-0}
  {\bibfield  {journal} {\bibinfo  {journal} {Nature Communications}\ }\textbf
  {\bibinfo {volume} {12}},\ \bibinfo {pages} {4034} (\bibinfo {year}
  {2021})}\BibitemShut {NoStop}%
\bibitem [{\citenamefont {Schiros}\ \emph {et~al.}(2012)\citenamefont
  {Schiros}, \citenamefont {Nordlund}, \citenamefont {Pálová}, \citenamefont
  {Prezzi}, \citenamefont {Zhao}, \citenamefont {Kim}, \citenamefont
  {Wurstbauer}, \citenamefont {Gutiérrez}, \citenamefont {Delongchamp},
  \citenamefont {Jaye}, \citenamefont {Fischer}, \citenamefont {Ogasawara},
  \citenamefont {Pettersson}, \citenamefont {Reichman}, \citenamefont {Kim},
  \citenamefont {Hybertsen},\ and\ \citenamefont
  {Pasupathy}}]{schiros2012connecting}%
  \BibitemOpen
  \bibfield  {author} {\bibinfo {author} {\bibfnamefont {Theanne}\ \bibnamefont
  {Schiros}}, \bibinfo {author} {\bibfnamefont {Dennis}\ \bibnamefont
  {Nordlund}}, \bibinfo {author} {\bibfnamefont {Lucia}\ \bibnamefont
  {Pálová}}, \bibinfo {author} {\bibfnamefont {Deborah}\ \bibnamefont
  {Prezzi}}, \bibinfo {author} {\bibfnamefont {Liuyan}\ \bibnamefont {Zhao}},
  \bibinfo {author} {\bibfnamefont {Keun~Soo}\ \bibnamefont {Kim}}, \bibinfo
  {author} {\bibfnamefont {Ulrich}\ \bibnamefont {Wurstbauer}}, \bibinfo
  {author} {\bibfnamefont {Christopher}\ \bibnamefont {Gutiérrez}}, \bibinfo
  {author} {\bibfnamefont {Dean}\ \bibnamefont {Delongchamp}}, \bibinfo
  {author} {\bibfnamefont {Cherno}\ \bibnamefont {Jaye}}, \bibinfo {author}
  {\bibfnamefont {Daniel}\ \bibnamefont {Fischer}}, \bibinfo {author}
  {\bibfnamefont {Hirohito}\ \bibnamefont {Ogasawara}}, \bibinfo {author}
  {\bibfnamefont {Lars G.~M.}\ \bibnamefont {Pettersson}}, \bibinfo {author}
  {\bibfnamefont {David~R.}\ \bibnamefont {Reichman}}, \bibinfo {author}
  {\bibfnamefont {Philip}\ \bibnamefont {Kim}}, \bibinfo {author}
  {\bibfnamefont {Mark~S.}\ \bibnamefont {Hybertsen}}, \ and\ \bibinfo {author}
  {\bibfnamefont {Abhay~N.}\ \bibnamefont {Pasupathy}},\ }\bibfield  {title}
  {\enquote {\bibinfo {title} {{Connecting Dopant Bond Type with Electronic
  Structure in N-Doped Graphene}},}\ }\href {\doibase 10.1021/nl301409h}
  {\bibfield  {journal} {\bibinfo  {journal} {Nano Letters}\ }\textbf {\bibinfo
  {volume} {12}},\ \bibinfo {pages} {4025--4031} (\bibinfo {year}
  {2012})}\BibitemShut {NoStop}%
\bibitem [{\citenamefont {Lippitz}\ \emph {et~al.}(2013)\citenamefont
  {Lippitz}, \citenamefont {Friedrich},\ and\ \citenamefont
  {Unger}}]{lippitz2013plasma}%
  \BibitemOpen
  \bibfield  {author} {\bibinfo {author} {\bibfnamefont {Andreas}\ \bibnamefont
  {Lippitz}}, \bibinfo {author} {\bibfnamefont {Jörg~F.}\ \bibnamefont
  {Friedrich}}, \ and\ \bibinfo {author} {\bibfnamefont {Wolfgang~E.S.}\
  \bibnamefont {Unger}},\ }\bibfield  {title} {\enquote {\bibinfo {title}
  {{Plasma bromination of HOPG surfaces: A NEXAFS and synchrotron XPS
  study}},}\ }\href {\doibase https://doi.org/10.1016/j.susc.2013.01.020}
  {\bibfield  {journal} {\bibinfo  {journal} {Surface Science}\ }\textbf
  {\bibinfo {volume} {611}},\ \bibinfo {pages} {L1--L7} (\bibinfo {year}
  {2013})}\BibitemShut {NoStop}%
\bibitem [{\citenamefont {Cody}(2000)}]{cody2000probing}%
  \BibitemOpen
  \bibfield  {author} {\bibinfo {author} {\bibfnamefont {George~D.}\
  \bibnamefont {Cody}},\ }\bibfield  {title} {\enquote {\bibinfo {title}
  {{Probing chemistry within the membrane structure of wood with soft X-ray
  spectral microscopy}},}\ \ }(\bibinfo {year} {2000})\ pp.\ \bibinfo {pages}
  {307--312}\BibitemShut {NoStop}%
\bibitem [{\citenamefont {Karunakaran}\ \emph {et~al.}(2015)\citenamefont
  {Karunakaran}, \citenamefont {Christensen}, \citenamefont {Gaillard},
  \citenamefont {Lahlali}, \citenamefont {Blair}, \citenamefont {Perumal},
  \citenamefont {Miller},\ and\ \citenamefont
  {Hitchcock}}]{karunakaran2015introduction}%
  \BibitemOpen
  \bibfield  {author} {\bibinfo {author} {\bibfnamefont {Chithra}\ \bibnamefont
  {Karunakaran}}, \bibinfo {author} {\bibfnamefont {Colleen~R}\ \bibnamefont
  {Christensen}}, \bibinfo {author} {\bibfnamefont {Cedric}\ \bibnamefont
  {Gaillard}}, \bibinfo {author} {\bibfnamefont {Rachid}\ \bibnamefont
  {Lahlali}}, \bibinfo {author} {\bibfnamefont {Lisa~M}\ \bibnamefont {Blair}},
  \bibinfo {author} {\bibfnamefont {Vijayan}\ \bibnamefont {Perumal}}, \bibinfo
  {author} {\bibfnamefont {Shea~S}\ \bibnamefont {Miller}}, \ and\ \bibinfo
  {author} {\bibfnamefont {Adam~P}\ \bibnamefont {Hitchcock}},\ }\bibfield
  {title} {\enquote {\bibinfo {title} {{Introduction of soft X-ray
  spectromicroscopy as an advanced technique for plant biopolymers
  research}},}\ }\href {\doibase 10.1371/journal.pone.0122959} {\bibfield
  {journal} {\bibinfo  {journal} {PloS one}\ }\textbf {\bibinfo {volume}
  {10}},\ \bibinfo {pages} {e0122959} (\bibinfo {year} {2015})}\BibitemShut
  {NoStop}%
\bibitem [{Note2()}]{Note2}%
  \BibitemOpen
  \bibinfo {note} {The simulations of the XANES spectra of nCL\protect \tmspace
  +\thinmuskip {.1667em} were performed by taken into account the polar angle
  [$\theta $, presented in Figs.\ref {xas-full-all} and \ref {xas-full-all2}],
  and two azimuthal ($\phi $) angles, one for polarization vectors parallel and
  another perpendicular to the cellulose fibrils.}\BibitemShut {Stop}%
\bibitem [{\citenamefont {Bader}(1990)}]{bader}%
  \BibitemOpen
  \bibfield  {author} {\bibinfo {author} {\bibfnamefont {R.}~\bibnamefont
  {Bader}},\ }\href@noop {} {\emph {\bibinfo {title} {Atoms in Molecules: A
  Quantum Theory}}}\ (\bibinfo  {publisher} {Oxford University Press},\
  \bibinfo {address} {New York},\ \bibinfo {year} {1990})\BibitemShut {NoStop}%
\bibitem [{Note3()}]{Note3}%
  \BibitemOpen
  \bibinfo {note} {The work function, $\Phi $, is defined as the energy
  position of the Fermi level with respect to the vacuum level, $\Phi
  =E_{\protect \rm F}-E_{\protect \rm vac}$, where $E_{\protect \rm vac}$ is
  obtained from the electrostatic potential calculation in a vacuum region far
  away from the system. Here we set $E_{\protect \rm vac}=0$.}\BibitemShut
  {Stop}%
\bibitem [{Note4()}]{Note4}%
  \BibitemOpen
  \bibinfo {note} {For each value of electric field, the atomic positions of
  the nCL$^{\protect \rm phob}$/G system were fully relaxed.}\BibitemShut
  {Stop}%
\bibitem [{\citenamefont {Si}\ \emph {et~al.}(2016)\citenamefont {Si},
  \citenamefont {Sun},\ and\ \citenamefont {Liu}}]{si2016strain}%
  \BibitemOpen
  \bibfield  {author} {\bibinfo {author} {\bibfnamefont {Chen}\ \bibnamefont
  {Si}}, \bibinfo {author} {\bibfnamefont {Zhimei}\ \bibnamefont {Sun}}, \ and\
  \bibinfo {author} {\bibfnamefont {Feng}\ \bibnamefont {Liu}},\ }\bibfield
  {title} {\enquote {\bibinfo {title} {{Strain engineering of graphene: a
  review}},}\ }\href {\doibase 10.1039/C5NR07755A} {\bibfield  {journal}
  {\bibinfo  {journal} {Nanoscale}\ }\textbf {\bibinfo {volume} {8}},\ \bibinfo
  {pages} {3207--3217} (\bibinfo {year} {2016})}\BibitemShut {NoStop}%
\bibitem [{\citenamefont {Miao}\ \emph {et~al.}(2021)\citenamefont {Miao},
  \citenamefont {Liang},\ and\ \citenamefont {Cheng}}]{miao2021straintronics}%
  \BibitemOpen
  \bibfield  {author} {\bibinfo {author} {\bibfnamefont {Feng}\ \bibnamefont
  {Miao}}, \bibinfo {author} {\bibfnamefont {Shi-Jun}\ \bibnamefont {Liang}}, \
  and\ \bibinfo {author} {\bibfnamefont {Bin}\ \bibnamefont {Cheng}},\
  }\bibfield  {title} {\enquote {\bibinfo {title} {{Straintronics with van der
  Waals materials}},}\ }\href {\doibase 10.1038/s41535-021-00360-3} {\bibfield
  {journal} {\bibinfo  {journal} {npj Quantum Materials}\ }\textbf {\bibinfo
  {volume} {6}},\ \bibinfo {pages} {1--4} (\bibinfo {year} {2021})}\BibitemShut
  {NoStop}%
\bibitem [{\citenamefont {Yankowitz}\ \emph {et~al.}(2016)\citenamefont
  {Yankowitz}, \citenamefont {Watanabe}, \citenamefont {Taniguchi},
  \citenamefont {San-Jose},\ and\ \citenamefont
  {LeRoy}}]{yankowitz2016pressure}%
  \BibitemOpen
  \bibfield  {author} {\bibinfo {author} {\bibfnamefont {Matthew}\ \bibnamefont
  {Yankowitz}}, \bibinfo {author} {\bibfnamefont {K}~\bibnamefont {Watanabe}},
  \bibinfo {author} {\bibfnamefont {T}~\bibnamefont {Taniguchi}}, \bibinfo
  {author} {\bibfnamefont {Pablo}\ \bibnamefont {San-Jose}}, \ and\ \bibinfo
  {author} {\bibfnamefont {Brian~J}\ \bibnamefont {LeRoy}},\ }\bibfield
  {title} {\enquote {\bibinfo {title} {{Pressure-induced commensurate stacking
  of graphene on boron nitride}},}\ }\href {\doibase 10.1038/ncomms13168}
  {\bibfield  {journal} {\bibinfo  {journal} {Nature communications}\ }\textbf
  {\bibinfo {volume} {7}},\ \bibinfo {pages} {1--8} (\bibinfo {year}
  {2016})}\BibitemShut {NoStop}%
\bibitem [{\citenamefont {Vincent}\ \emph {et~al.}(2018)\citenamefont
  {Vincent}, \citenamefont {Panchal}, \citenamefont {Booth}, \citenamefont
  {Power}, \citenamefont {Jauho}, \citenamefont {Antonov},\ and\ \citenamefont
  {Kazakova}}]{vincent2018probing}%
  \BibitemOpen
  \bibfield  {author} {\bibinfo {author} {\bibfnamefont {Tom}\ \bibnamefont
  {Vincent}}, \bibinfo {author} {\bibfnamefont {Vishal}\ \bibnamefont
  {Panchal}}, \bibinfo {author} {\bibfnamefont {Tim}\ \bibnamefont {Booth}},
  \bibinfo {author} {\bibfnamefont {Stephen~R}\ \bibnamefont {Power}}, \bibinfo
  {author} {\bibfnamefont {Antti-Pekka}\ \bibnamefont {Jauho}}, \bibinfo
  {author} {\bibfnamefont {Vladimir}\ \bibnamefont {Antonov}}, \ and\ \bibinfo
  {author} {\bibfnamefont {Olga}\ \bibnamefont {Kazakova}},\ }\bibfield
  {title} {\enquote {\bibinfo {title} {{Probing the nanoscale origin of strain
  and doping in graphene-hBN heterostructures}},}\ }\href {\doibase
  10.1088/2053-1583/aaf1dc} {\bibfield  {journal} {\bibinfo  {journal} {2D
  Materials}\ }\textbf {\bibinfo {volume} {6}},\ \bibinfo {pages} {015022}
  (\bibinfo {year} {2018})}\BibitemShut {NoStop}%
\bibitem [{\citenamefont {Forestier}\ \emph {et~al.}(2020)\citenamefont
  {Forestier}, \citenamefont {Balima}, \citenamefont {Bousige}, \citenamefont
  {Pinheiro}, \citenamefont {Fulcrand}, \citenamefont {Kalbáč}, \citenamefont
  {Machon},\ and\ \citenamefont {San-Miguel}}]{forestier2020strain}%
  \BibitemOpen
  \bibfield  {author} {\bibinfo {author} {\bibfnamefont {Alexis}\ \bibnamefont
  {Forestier}}, \bibinfo {author} {\bibfnamefont {Félix}\ \bibnamefont
  {Balima}}, \bibinfo {author} {\bibfnamefont {Colin}\ \bibnamefont {Bousige}},
  \bibinfo {author} {\bibfnamefont {Gardênia de~Sousa}\ \bibnamefont
  {Pinheiro}}, \bibinfo {author} {\bibfnamefont {Rémy}\ \bibnamefont
  {Fulcrand}}, \bibinfo {author} {\bibfnamefont {Martin}\ \bibnamefont
  {Kalbáč}}, \bibinfo {author} {\bibfnamefont {Denis}\ \bibnamefont
  {Machon}}, \ and\ \bibinfo {author} {\bibfnamefont {Alfonso}\ \bibnamefont
  {San-Miguel}},\ }\bibfield  {title} {\enquote {\bibinfo {title} {{Strain and
  Piezo-Doping Mismatch between Graphene Layers}},}\ }\href {\doibase
  10.1021/acs.jpcc.0c01898} {\bibfield  {journal} {\bibinfo  {journal} {The
  Journal of Physical Chemistry C}\ }\textbf {\bibinfo {volume} {124}},\
  \bibinfo {pages} {11193--11199} (\bibinfo {year} {2020})}\BibitemShut
  {NoStop}%
\end{thebibliography}%

\end{document}